\documentclass[aps,prd,onecolumn,groupedaddress,preprintnumbers,superscriptaddress,showpacs,nofootinbib]{revtex4}
\usepackage{amsmath,amssymb,latexsym}

\usepackage{graphicx}

\begin{document}

\title{
Instability of de Sitter brane and horizon entropy in a 6D braneworld 
}


\author{Shunichiro Kinoshita}
\email[]{kinoshita_at_utap.phys.s.u-tokyo.ac.jp}
\affiliation{   
Department of Physics, Graduate School of Science, The University of Tokyo,
Hongo 7-3-1, Bunkyo-ku, Tokyo 113-0033, Japan
}
\author{
Yuuiti Sendouda
}\email[]{sendouda_at_yukawa.kyoto-u.ac.jp}
\affiliation{
Yukawa Institute for Theoretical Physics, Kyoto University,
Kitashirakawa Oiwake-cho, Sakyo-ku, Kyoto 606-8502, Japan
}
\author{
Shinji Mukohyama
}\email[]{mukoyama_at_phys.s.u-tokyo.ac.jp}
\affiliation{
Department of Physics, Graduate School of Science, The University of Tokyo,
Hongo 7-3-1, Bunkyo-ku, Tokyo 113-0033, Japan
}
\affiliation{
Research Center for the Early Universe, Graduate School of Science, The University of Tokyo,
Hongo 7-3-1, Bunkyo-ku, Tokyo 113-0033, Japan
}

\preprint{UTAP-575} 
\preprint{RESCEU-2/07}

\date{\today}

\begin{abstract}
 We investigate thermodynamic and dynamical stability of a family of
 six-dimensional braneworld solutions with de Sitter branes. First, we
 investigate thermodynamic stability in terms of de Sitter entropy. We
 see that the family of solutions is divided into two distinct branches:
 the high-entropy branch and the low-entropy branch. By analogy with
 ordinary thermodynamics, the high-entropy branch is expected to be
 stable and the low-entropy branch to be unstable. Next, we investigate
 dynamical stability by analysing linear perturbations around the
 solutions. Perturbations are decomposed into scalar, vector and tensor
 sectors according to the representation of the 4D de Sitter symmetry,
 and each sector is analysed separately. It is found that when the
 Hubble expansion rates on the branes are too large, there appears a
 tachyonic mode in the scalar sector and the background solution becomes
 dynamically unstable. We show analytically that the onset of the
 thermodynamic instability and that of the dynamical instability exactly
 coincide. Therefore, the braneworld model provides a new example
 illustrating close relations between thermodynamic and dynamical
 instability.
\end{abstract}

\maketitle

\section{Introduction}

The braneworld scenario, in which matter fields are confined on a brane
in higher dimensional spacetime, has been one of the most exciting
subjects in the research into the early universe. As the early universe is
supposed to have experienced a rather high-energy epoch, studies of the 
dynamics of the early universe might reveal as yet undiscovered imprints of
extra dimensions to us.

In our previous
studies~\cite{Mukohyama:2005yw,Yoshiguchi:2005nn,Sendouda:2006bn} we
considered a $6$-dimensional braneworld model with one or two
$3$-branes. The shape and the size of the extra dimensions are
stabilized by magnetic flux of a $U(1)$ gauge field and the geometry on
each brane is either Minkowski or de Sitter. Since all moduli are
stabilized, it is expected that the $4$-dimensional Einstein gravity
should be recovered on each brane. In Ref.~\cite{Mukohyama:2005yw} we
analyzed how the Hubble expansion rate on each brane changes when the
brane tension changes. The resulting relation between the Hubble
expansion rate and the brane tension can be considered as an effective
Friedmann equation, which in turn defines an effective Newton's
constant. It was shown that at sufficiently low Hubble expansion rates,
the effective Newton's constant resulting from this analysis agrees with
that inferred by simply integrating extra dimensions out. The stability
of this model was investigated in
Ref.~\cite{Yoshiguchi:2005nn,Sendouda:2006bn}, where it was shown that
the background with Minkowski branes is always stable against linear
perturbations with axisymmetry in the extra dimensions.

The purpose of this paper is to study stability of backgrounds with de
Sitter branes. This is not just a trivial extension of the previous
analysis but actually results in a qualitatively different
picture. Indeed, the background becomes unstable when the Hubble
expansion rate becomes too large. 
(For other models, see for example~\cite{Frolov:2003yi,Contaldi:2004hr,Bousso:2002fi,Martin:2004wp,Krishnan:2005su}.)
We shall obtain the critical expansion
rate at which the background becomes unstable and interpret this
behavior in terms of de Sitter thermodynamics.

Spacetimes with horizons appear mysterious in many respects. An 
observer does not have causal contact with any events beyond a horizon
and, in this sense, the horizon separates the region behind it from the
observer's side. Thus, from the observer's viewpoint, information beyond
the horizon is completely irrelevant (as far as classical evolution in
the observable region is concerned). This, nothing but a restatement of
the definition of horizons, suggests a coarse-grained description of a 
physical system in or of a spacetime with horizons. It is therefore not
totally surprising that a spacetime with horizons exhibits properties
analogous to those of thermodynamics. Indeed, in the case of black holes
such properties are known as black hole thermodynamics~\footnote{For a
review see e.g.~\cite{Mukohyama:1998ng} and references therein.}.

Black hole entropy, defined as one quarter of horizon area in Planck
units, is central to black hole thermodynamics. The second law
of black hole thermodynamics states that black hole entropy never
decreases classically. Its semi-classical extension, known 
as the generalized second law, states that the sum of black hole entropy
and matter entropy does not decrease. These second laws are direct
analogues of the second law of ordinary thermodynamics and restrict
directions in which a physical system involving black holes can 
evolve. In ordinary thermodynamics it is due to the second law that 
entropy can be used to analyze stability of a thermodynamic system.
Therefore, the validity of the second law of black hole
thermodynamics suggests potential use of black hole entropy to guess 
stability of a system including black holes or objects (branes, strings,
rings, and so on)~\cite{Braden:1990hw,Maeda:1993ap,Prestidge:1999uq,Gubser:2000mm,Gubser:2000ec,Reall:2001ag,Arcioni:2004ww}. 
(Also, for a review see e.g.~\cite{Harmark:2007md} and references therein.)

The close connection between black hole event horizons and thermodynamics
can be extended to event horizons in cosmological models with a positive
cosmological constant~\cite{Gibbons:1977mu}. The de Sitter
thermodynamics includes concepts of de Sitter temperature and de Sitter
entropy.

In this paper we shall extend the de Sitter entropy to our braneworld
set-up with two extra dimensions and use it to ``predict'' the stability
condition for the backgrounds with de Sitter branes. Intriguingly, the
simple ``prediction'' by de Sitter entropy completely agrees with the
result of rather involved analysis of linear perturbations. This is
really non-trivial and may be considered as an evidence showing the
usefulness of de Sitter entropy and, more generally, holographic 
principle~\cite{Strominger:2001pn, Bousso:2002ju}.

The rest of this paper is organized as follows. In Sec.~\ref{sec:6DWFC} 
we briefly review our $6$-dimensional braneworld model with de Sitter branes.
In Sec.~\ref{sec:thermodynamics} we study thermodynamic stability of the
spacetime by using entropy.
In Sec.~\ref{sec:dynamical}, in order to investigate dynamical stability
 we consider linear perturbation around the background solution.
In Sec.~\ref{sec:summary} we summarize the results and discuss them.

\section{6D brane model with warped flux compactification}
\label{sec:6DWFC}

We consider a 6D Einstein--Maxwell system described by the action 
\begin{equation}
 I = \frac{1}{16\pi G_6}\int \mathrm d^6x \sqrt{-g} \left(R - 2\Lambda_6 - \frac{1}{2}F_{MN}F^{MN}\right),
\end{equation}
where $\Lambda_6$ is the 6D bulk cosmological constant $(\Lambda_6 > 0)$ 
and $F_{MN} = \partial_M A_N - \partial_N A_M$ is the field strength of
the $U(1)$ gauge field $A_M$ introduced to stabilize the extra
dimensions. (We use units with $G_6 = 1$ where $G_6$ is the 6D Newton's 
constant unless otherwise noted.) The braneworld solution is 
\begin{equation}
 \mathrm ds^2_6 = \tilde r^2 \mathrm ds^2_4 
  + \frac{\mathrm d\tilde r^2}{\tilde f(\tilde r)} + \tilde f(\tilde r)\mathrm d\tilde\phi^2,\quad 
  A_M\mathrm dx^M = \tilde A(\tilde r)\mathrm d\tilde\phi,
\end{equation}
where
\begin{equation}
  \tilde f(\tilde r) = k - \frac{\Lambda_6}{10}\tilde r^2 -
   \frac{\tilde{\mathcal M}}{\tilde r^3} -
   \frac{\tilde b^2}{12\tilde r^6},\quad
  \tilde A(\tilde r) = \frac{\tilde b}{3\tilde r^3},
\end{equation} 
and $\mathrm ds^2_4$ is the metric of the 4-dimensional spacetime with a
constant curvature $k$, i.e.\ de Sitter ($k>0$), Minkowski ($k=0$), or
anti-de Sitter (AdS) ($k<0$) spacetime. Without loss of generality, we
can normalize $k$ to $\pm 1$ except for the Minkowski ($k=0$) case. We 
consider the range of parameters in which $\tilde f(\tilde r)$ has two positive roots,
and locate two 3-branes at the positions of the two roots $\tilde r=r_\pm$ 
$(0 < r_- < r_+)$. We call the branes at $\tilde r=r_{\pm}$ the $\tilde r_{\pm}$-branes,
respectively. The tension of each brane determines a conical deficit in
the extra dimensions. 
To be precise,
the period of the angular coordinate ($\tilde\phi \sim \tilde\phi + \Delta\phi$) is
given by 
\begin{equation}
 \Delta\phi = \frac{2\pi - 8\pi\sigma_+}{|\tilde f'(r_+)/2|}
  = \frac{2\pi - 8\pi\sigma_-}{|\tilde f'(r_-)/2|},\label{eq:Delta_phi}
\end{equation}
where $\sigma_\pm$ are tensions of the branes at $\tilde r=r_\pm$.

This braneworld model includes two compact extra dimensions and two
3-branes at $\tilde r=r_\pm$ with the tensions $\sigma_\pm$. For $k=+1$, the
geometries on the brane at $\tilde r=r_\pm$ are the 4D de Sitter
spacetimes with the Hubble expansion rates $h_\pm = 1/r_\pm$,
respectively.
For $k=0$, the brane geometry is the Minkowski spacetime. 
In each case the energy scale on each brane is scaled by the warp factor
$\tilde r^2$ depending on the position of the brane. The role of the
$U(1)$ field is to stabilize extra dimensions with magnetic flux.

Now, for later convenience we re-scale coordinates and parameters of the
geometry with the 4D Hubble parameter $h_+$ on the $r_+$-brane in the
following manner: 
\begin{equation}
 \tilde r \to \frac{r}{h}, \quad \tilde\phi \to h\phi, 
  \quad \tilde{\mathcal M} \to \frac{\mathcal M}{h^5},
  \quad \tilde b \to \frac{b}{h^4}, \quad \tilde x^\mu \to h x^\mu,
\end{equation}
where we have omitted the subscript ``$+$'' of $h_+$ and the tilde denotes
re-scaled quantities. After this re-scaling, we obtain 
\begin{equation}
  \mathrm ds^2_6 = r^2 g_{\mu\nu}
   \mathrm d x^\mu\mathrm d x^\nu 
   + \frac{\mathrm d r^2}{f(r)}
   + f(r)\mathrm d\phi^2,\quad 
  A_M\mathrm dx^M = A(r)\mathrm d\phi,
  \label{eq:background1}
\end{equation}
where
\begin{equation}
  f(r) = h^2 - \frac{\Lambda_6}{10}r^2
   - \frac{\mathcal M}{r^3} - \frac{b^2}{12r^6},\quad
  A(r) = \frac{b}{3r^3}.
 \label{eq:background2}
\end{equation}
Here $g_{\mu\nu}$ is the metric of the 4D de Sitter spacetime with the
Hubble parameter $h$. 
The metric function now satisfies $f(\alpha) = f(1) = 0$,
where $\alpha \equiv r_-/r_+$ is the ratio of the warp factors at the
branes and, at the same time, describes how the geometry of the 2D extra
dimensions is warped. 
Note that in the special case $\alpha = 1$ the geometry is not warped but
locally a round sphere.

Advantages of this re-scaled parameterization are as follows. First,
the new parameters and coordinates remain finite and regular in the  
limit where the geometries on the branes become Minkowski. Second,
since the warp factor evaluated at the $r_+$-brane is unity 
($r=1$), many quantities rooted in these coordinates have
physical meaning as observables measured by observers on the
$r_+$-brane. As we will see later, a Kaluza--Klein (KK) mass
measured on the $r_+$-brane remains finite in the $\alpha \to 0$ limit,
while that on the $r_-$-brane diverges. Hence, it is useful (in
particular for the purpose of numerical calculations) to use quantities
measured on the $r_+$-brane. Of course, if we know the value of a
quantity measured on the $r_+$ brane then we can easily obtain the
corresponding value measured on the $r_-$-brane by multiplying the warp factor
$\alpha$, e.g., $h_- = h_+/\alpha$, $m^2_- = m^2_+/\alpha^2$, and so on,
where $m_\pm$ is the KK mass observed on the $r_\pm$-brane.

We introduce another convenient parameter $\beta$. From
$f(\alpha) = f(1) = 0$ we obtain the relation 
\begin{equation}
 h^2 + \frac{b^2}{12\alpha^3} =
  \frac{\Lambda_6 (\alpha^4 + \alpha^3 + \alpha^2 + \alpha + 1)}
   {10 (\alpha^2 + \alpha + 1)}.
\end{equation}
Since $b^2 \ge 0$, the Hubble parameter $h$ has a maximum value
$h_\mathrm{max}(\alpha)$ for a given $\alpha$. Hence we define $\beta$ by 
\begin{equation}
 \beta \equiv \frac{h^2}{h_\mathrm{max}(\alpha)^2}
  = \frac{10(1 + \alpha + \alpha^2)}
  {\Lambda_6(1 + \alpha + \alpha^2 + \alpha^3 + \alpha^4)}h^2,
\end{equation}
which is the squared Hubble parameter $h^2$ on the $r_+$-brane
normalized by $h_\mathrm{max}(\alpha)$. Alternatively,
$\beta$ could be defined as $h_-^2$ normalized by its maximum value, but
the two definitions are actually equivalent. Thus, $\beta$ describes how
much the $4$D part of the metric is curved. When
$\beta=0$, the 4D spacetime becomes flat~\footnote{In fact, if  
$\beta < 0$ then the solution represents a braneworld with AdS
branes. In this paper, however, we shall not consider this case,
focusing only on the de Sitter ($0<\beta\leq 1$) and Minkowski
($\beta=0$) branes.}. 
In contrast, in the maximal case ($\beta = 1$), the flux disappears:
$b=0$.

For a fixed bulk cosmological constant $\Lambda_6$ and a fixed
discrete parameter $k$ ($=0,1$), the solution is locally parameterized
by the two parameters $b$ and $\mathcal M$. However it is more convenient to use
the pair of dimension-less parameters $(\alpha,\beta)$, both of which
run over the finite interval $[0,1]$. This parameterization includes
both $k=1$ ($0<\beta\leq 1$) and $k=0$ ($\beta=0$) cases, and treats the
de Sitter and the Minkowski branes in the common way.

In the following sections we use the parameters ($\alpha$, $\beta$) and the
coordinates ($r$, $\phi$, $x^{\mu}$) to describe
the background geometry.

\section{thermodynamic stability}
\label{sec:thermodynamics}

As stated in the introduction, one of the main subjects in this paper is
the close connection between thermodynamic and dynamical properties of
spacetimes. In particular, we shall see exact agreement between
thermodynamic and dynamical stabilities for the $6$-dimensional brane
world solutions described in the previous section. In this section we
discuss thermodynamic stability by using the de Sitter entropy. We also
point out the close relation between the specific heat and the effective
Friedmann equation.

\subsection{Entropy argument}

In the $4$-dimensional Einstein gravity, a de Sitter space has
entropy given by one quarter of the area of the
horizon~\cite{Gibbons:1977mu}. This is an analogue of the well-known
black hole entropy and has been the basis of holographic arguments for 
de Sitter spaces~\cite{Strominger:2001pn, Bousso:2002ju}.

In the 6D braneworld background solution, each point in the 2D extra
dimensions corresponds to a 4D de Sitter space. Therefore, we shall
define the de Sitter entropy in our context by one quarter of the area of
the cosmological horizon integrated over the extra dimensions, i.e., one
quarter of the volume of the 4-dimensional surface foliated by de Sitter
horizons: 
\begin{equation}
 S \equiv \frac{\mathcal A_4}{4 G_6},
\end{equation}
where $\mathcal A_4$ is the 4-dimensional volume given by
\begin{equation}
 \mathcal A_4 = \frac{4\pi}{h^2} \int_\alpha^1 \!\! r^2 \mathrm dr
  \int^{\Delta\phi}_0\!\!\mathrm d\phi
  = \frac{4\pi}{3h^2}(1 - \alpha^3)\Delta\phi,
\end{equation}
and $G_6$ is the 6D Newton's constant. 
Note that, for the convenience of the discussion, we have temporarily
restored $G_6$ in this paragraph.
Actually, it is easy to show that
this definition agrees with the definition of the 4D de Sitter entropy
that observers on each brane would adopt. The area of the de Sitter
horizon and the Newton's constant on the $r_{\pm}$-brane are,
respectively, given by 
\begin{equation}
 \mathcal A_+ = \alpha^{-2}\mathcal A_- = \frac{4\pi}{h^2}
\end{equation}
and 
\begin{equation}
 \frac{1}{G_{\mathrm N +}} = \frac{\alpha^2}{G_{\mathrm N -}} = 
  \frac{1}{G_6} \int_\alpha^1 \!\! r^2 \mathrm dr
  \int^{\Delta\phi}_0\!\!\mathrm d\phi.\label{eqn:GN}
\end{equation}
Thus, the 4D de Sitter entropies on the $r_{\pm}$-branes are 
\begin{equation}
 S_{\pm} \equiv \frac{\mathcal A_\pm}{4G_{\mathrm N \pm}},
\end{equation}
respectively. These do indeed agree with the above definition based on the 
horizon area integrated over the extra dimensions:
\begin{equation}
 S_+ = S_- = S.
\end{equation}
In the 4D picture, the integration over extra dimensions is taken care
of by the above formula of the 4D Newton's constant (\ref{eqn:GN}). 
Since the three different definitions agree, we have a unique definition
of de Sitter entropy in our braneworld set-up.

To investigate the thermodynamic stability of the system, it is useful
to consider the de Sitter entropy as a function of conserved
quantities. The family of solutions described in the previous section 
has three conserved quantities: the total magnetic flux and the
tensions of two branes. The total magnetic flux $\Phi$ is given by 
\begin{equation}
 \Phi = \int_\alpha^1 F_{r\phi} \mathrm dr \int_0^{\Delta\phi} \mathrm d\phi =
  \frac{b}{3\alpha^3}(\alpha^3 - 1)\Delta\phi. 
\end{equation}
By Eq.~(\ref{eq:Delta_phi}), tensions $\sigma_{\pm}$ of the
$r_{\pm}$-branes are expressed as 
\begin{equation}
 \sigma_\pm = \frac{2\pi-\kappa_\pm}{8\pi},
\end{equation}
respectively, where $\kappa_{\pm}$ are defined by
\begin{equation}
 \kappa_+ = -\frac{1}{2}f'(1)\Delta\phi, \quad
  \kappa_- = \frac{1}{2}f'(\alpha)\Delta\phi.
\end{equation}
Since there is one-to-one correspondence between $\sigma_{\pm}$ and
$\kappa_{\pm}$, we can consider the de Sitter entropy $S$ as a function
of either ($\Phi$, $\sigma_+$, $\sigma_-$) or ($\Phi$, $\kappa_+$,
$\kappa_-$).

Variables $S$, $\Phi$ and $\kappa_\pm$ defined above are proportional 
to $\Delta\phi$. Because of this trivial dependence, we can eliminate 
$\Delta\phi$ from all thermodynamic considerations by properly
normalizing the variables. In particular, in 
Appendix~\ref{sec:thermodynamic_relatios} it is shown that the entropy
normalized by $\kappa_-$, $\widehat S \equiv S/\kappa_-$, satisfies the
differential relation 
\begin{equation}
 \mathrm d \widehat S = \left(\frac{\pi}{3h^4}\right) \mathrm d\eta +
  \left(\frac{\pi b}{3h^4}\right)\mathrm d\widehat\Phi,
  \label{eq:first_law}
\end{equation}
where $\eta\equiv \kappa_+ / \kappa_-$, and 
$\widehat\Phi \equiv \Phi / \kappa_-$ is the normalized total flux. This
means that the normalized entropy $\widehat S$ is a function of the two
variables $\eta$ and $\widehat\Phi$: $\widehat S (\eta, \widehat\Phi)$. 
Hereafter, $\eta$ and $\widehat\Phi$ are to be considered as two
independent thermodynamic variables. This accords with the observation
made in the previous section: the family of solutions are parameterized
by two parameters ($\alpha$, $\beta$) after proper scaling of 
variables. The angular period $\Delta\phi$ has been eliminated so that
$\widehat S$, $\eta$ and $\widehat\Phi$ do not depend on it.

Note that the normalization factor $1/\kappa_-$ is just a fixed
constant from the viewpoint of observers on the $r_+$-brane since these
observers are to see the response of the 4D geometry induced on the
$r_+$-brane to the change of the tension $\sigma_+$ or equivalently,
$\kappa_+$. In particular, with $\kappa_-$ fixed, we have 
\begin{equation}
 \mathrm d\eta = - \frac{8\pi}{\kappa_-}\mathrm d\sigma_+.
  \label{eqn:deta}
\end{equation}
Thus, the relation (\ref{eq:first_law}) is relevant for observers on the
$r_+$-brane (but not for those on the $r_-$-brane). Alternatively, we
could write down a differential relation relevant for observers on the
$r_-$-brane by normalizing the variables by the factor $1/\kappa_+$,
which is a constant from the viewpoint of observers on the
$r_-$-brane. In this sense, we have two conceptually different pictures:
one from $r_+$-brane perspectives and the other from $r_-$-brane
perspectives. The result of the thermodynamic stability analysis does
not depend on the picture: the result from one picture completely agrees
with that from the other picture. In the following, we shall see the
thermodynamic stability from the $r_+$-brane perspectives. Thus, unless
there is a possibility of confusion, we omit the subscripts ``$\pm$''.

\begin{figure}[t]
 \begin{center}
  \includegraphics[width=.6\linewidth]{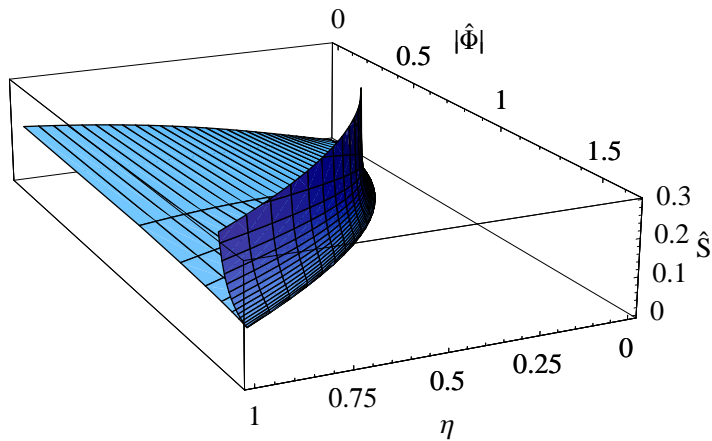}
  \caption{The normalized entropy $\widehat S$ as a function of $\eta$
  and $\widehat\Phi$. Different points on the surface 
  $\widehat S=\widehat S(\eta,\widehat\Phi)$ represent different
  background solutions, and there is no background solution outside the
  domain of $\widehat S(\eta,\widehat\Phi)$. It is easily seen that
  $\widehat S(\eta,\widehat\Phi)$ is not single-valued but actually
  double-valued in some part of its domain. Thus, the family of
  background solutions has two branches, divided by a  critical curve on
  which some derivatives of $\widehat S(\eta,\widehat\Phi)$ are
  singular. The high-entropy branch is thermodynamically preferred to
  the low-entropy branch. }
  \label{fig:entropy}
 \end{center}
\end{figure}
Fig.~\ref{fig:entropy} shows the graph of the normalized de Sitter
entropy $\widehat S$ as a function of the conserved quantities $\eta$
and $\widehat\Phi$. As we can easily see, 
$\widehat S(\eta,\widehat\Phi)$ is not single-valued but actually
double-valued in some part of its domain. It is also easy to see that
derivatives of $\widehat S(\eta,\widehat\Phi)$ are singular on a part of
the boundary of the domain. We call this part of the boundary of the
domain the {\it critical curve}. Different points on the surface 
$\widehat S=\widehat S(\eta,\widehat\Phi)$ represent different
background solutions, i.e.\ different spacetimes. (There is no background
solution outside the domain of $\widehat S(\eta,\widehat\Phi)$.)
Therefore, the critical curve divides the family of solutions into two
distinct branches: one with larger $\widehat S$ and the other with
smaller $\widehat S$. We call them the {\it high-entropy branch} and the
{\it low-entropy branch}, respectively. By definition, these two
branches merge on the critical curve. Note that in some region of the
low-entropy branch, $\widehat S(\eta,\widehat\Phi)$ is single-valued.

If the de Sitter entropy has any physical meaning like entropy in
ordinary thermodynamics then the high-entropy branch must be
thermodynamically preferred in some sense and must be more stable than
the low-entropy branch. The spacetime is marginally stable on the
critical curve. These observations are based only on our intuition that
the de Sitter entropy should play a role similar to that of entropy in
ordinary thermodynamics. 
Nonetheless, surprisingly, we shall see in the next
section that these observations correctly ``predict'' the result of
rather detailed analysis of dynamical stability against linear
perturbations. 

We now derive an explicit equation describing the critical curve. For
this purpose, we consider the coordinate transformation from 
$(\alpha, \beta)$ defined in the previous section to 
$(\eta, \widehat\Phi)$ defined above. As we shall see below, the
critical curve is the set of points where this coordinate transformation
becomes singular. As noticed in Sec.~\ref{sec:6DWFC}, $\alpha$ 
($\in (0,1]$)is the ratio of the warp factors of two branes and
describes the shape of the two-dimensional extra space. On the other 
hand $\beta$ ($\in [0,1]$) represents the curvature of 4D de Sitter
space normalized by the maximum value
$h^2_\mathrm{max}(\alpha)$ for a given $\alpha$. Since the background
solution is uniquely specified by $(\alpha, \beta)$ up to the scaling
explained in the previous section, any (properly normalized) quantities
can be rewritten as functions of $(\alpha, \beta)$. In particular, we
have two single-valued functions $\eta(\alpha, \beta)$ and 
$\widehat \Phi(\alpha,\beta)$ as a map 
$(\alpha, \beta) \to (\eta, \widehat\Phi)$. 
The entropy $\widehat S$ is also a single-valued function of 
$(\alpha, \beta)$ and, thus, the critical curve is determined by
vanishing Jacobian determinant of this map:
\begin{equation}
 \left|\frac{\partial (\eta, \widehat\Phi)}{\partial (\alpha, \beta)}\right| =
  0.\label{eqn:critical-curve-Jacobian}
\end{equation}
This condition can be solved with respect to $\beta$ to give the
following expression for the critical value of $\beta$:
\begin{equation}
 \begin{aligned}
 \beta_\mathrm{cri}(\alpha)
  &=
  \frac{1}
       {12(1+\alpha+\alpha^2)^2 (1+\alpha+\alpha^2+\alpha^3+\alpha^4)}
  \bigl[11 + 33\alpha + 66\alpha^2 + 85\alpha^3 + 90\alpha^4 + 85\alpha^5 +
   66\alpha^6 + 33\alpha^7 +  11\alpha^8 \\
  &\quad - (1 + \alpha) (1 + 2\alpha + 4\alpha^2 + 2\alpha^3 + \alpha^4)
     \sqrt{1 + 34\alpha^3 + \alpha^6}\bigr].
 \end{aligned}
 \label{eq:critical_curve}
\end{equation}
This is the analytic expression for the critical curve. 
As we see later, $\beta_\mathrm{cri}(1) = 2/3$ (or, 
$h^2_\mathrm{cri} = \Lambda_6 /9$ and $b^2_\mathrm{cri} = 6h^2_\mathrm{cri}$)
is nothing but the critical value for stabilizing the extra dimensions like
in Freund--Rubin compactification~\cite{Martin:2004wp}.

\subsection{Effective Newton's constant}

We have seen that the (normalized) de Sitter entropy is not a 
single-valued function but actually a double-valued function of the conserved
quantities $(\eta, \widehat\Phi)$ in some region of the domain. This
observation has led us to a rather natural stability criterion that
the high-entropy branch should be stable and that the low-entropy branch
should be unstable. This is a global statement: we would not be
able to say one of the two branches should be stable or unstable without
knowing the existence of the other branch.

In this subsection, we look at the stability from a different
viewpoint, considering the effective Friedmann equation investigated in
Ref.~\cite{Mukohyama:2005yw}. In particular, we shall see that the
thermodynamic stability is equivalent to the positivity of an
{\it effective Newton's constant}. Note that, while the thermodynamic
stability is a global statement, the positivity of the effective
Newton's constant is a local statement.

In our previous work~\cite{Mukohyama:2005yw} we showed that the 4D
Friedmann equation is recovered on the brane at sufficiently low energy
as the response of the 4D Hubble expansion rate to the change of the
brane tension. To be more precise, for small Hubble
expansion rates $h$, $h^2$ is expanded as 
\begin{equation}
 h^2 = \left.\frac{8\pi G_\mathrm N}{3}\right|_{\sigma=\sigma_0}
  (\sigma - \sigma_0) + \mathcal
  O((\sigma - \sigma_0)^2),
\end{equation}
where $\sigma$ is the brane tension and the 4D Newton's constant
$G_\mathrm N$ is obtained by simply integrating extra dimensions out as 
(\ref{eqn:GN}). The constant $\sigma_0$ is the value of $\sigma$ for the
Minkowski brane ($h=0$) and depends on $\Phi$ and $\sigma_-$, which are
actually constants from the viewpoint of observers on the
$r_+$-brane. Note that the 4D Newton's constant $G_\mathrm N$ defined as 
(\ref{eqn:GN}) is always positive.

\begin{figure}[t]
 \begin{center}
  \includegraphics[width=.6\linewidth]{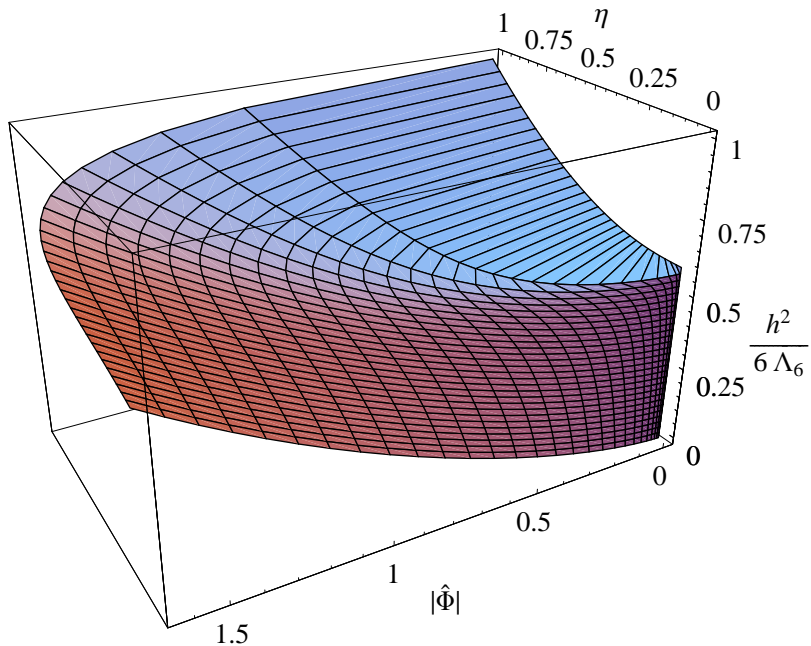}
  \caption{The Hubble expansion rate squared, $h^2$, as a function of
  $\eta$ and $\widehat\Phi$. For small $h^2$, 
  $(\partial h^2/\partial\eta)_{\widehat\Phi}$ is negative. However, as
  $h^2$ increases, the slope turns over and 
  $(\partial h^2/\partial\eta)_{\widehat\Phi}$ becomes positive. 
  (This figure is identical to Fig.~3 of Ref.~\cite{Mukohyama:2005yw}.)}
  \label{fig:Hubble}
 \end{center}
\end{figure}

The terms of order $\mathcal O((\sigma - \sigma_0)^2)$ and higher are small
compared with the linear term if the Hubble expansion rate is
sufficiently lower than the energy scale set by the bulk cosmological
constant. However, for larger $h^2$, the higher order corrections become
relevant and $h^2$ is no longer approximately linear. Moreover, as we
can see from Fig.~\ref{fig:Hubble}, $h^2$ is not necessarily an
increasing function of the brane tension. Indeed, if $h^2$ is larger
than a critical value then it is a decreasing function of the brane
tension. To make this peculiar behavior more quantitative, let us define
the effective Newton's constant $G_\mathrm{eff}$ on the $r_+$-brane by 
\begin{equation}
 G_\mathrm{eff} = \frac{3}{8\pi}
  \left(\frac{\partial h^2}{\partial\sigma_+}\right)_{\Phi,\sigma_-},
\end{equation}
and characterize the critical value by $1/G_\mathrm{eff}=0$. For $h$ smaller
(or larger) than the critical value, $G_\mathrm{eff}$ is positive (or negative,
respectively).

Note that, since this effective Newton's constant runs as the Hubble
expansion rate $h$ changes, the exact agreement with $G_\mathrm N$ is
met only in the $h^2\to +0$ limit. The agreement in this limit itself is
a rather non-trivial consequence of the dynamics of bulk spacetime as
explicitly shown in \cite{Mukohyama:2005yw}, but physical interpretation
is simple. While $G_\mathrm N$ was determined via the coefficient of the
contribution of the graviton zero mode to the higher dimensional action, 
$G_\mathrm{eff}$ here incorporates contributions of all Kaluza--Klein modes as
well as the zero mode. Since Kaluza--Klein contributions are irrelevant
at low energies, these two definitions should agree. On the other hand,
at high energies the Kaluza--Klein modes become relevant and make
$G_\mathrm{eff}$ deviate from $G_\mathrm N$.

In order to examine the curve $1/G_\mathrm{eff}=0$, the following observation
is useful:
\begin{equation}
 G_\mathrm{eff} = - \frac{3}{\kappa_-}
  \left(\frac{\partial h^2}{\partial \eta}\right)_{\widehat\Phi}
  = - \frac{6h}{\kappa_-}
  \left|\frac{\partial(h,\widehat\Phi)}{\partial(\alpha,\beta)}\right|
  \left/\left|
	 \frac{\partial(\eta,\widehat\Phi)}{\partial(\alpha,\beta)}
	\right|\right. .
\end{equation}
Since $h$ and $\widehat\Phi$ are single-valued smooth functions of 
$(\alpha,\beta)$, the Jacobian 
$|\partial(h,\widehat\Phi)/\partial(\alpha,\beta)|$
does not diverge. Hence, $G_\mathrm{eff}$ diverges if and only if the
denominator 
$|\partial(\eta,\widehat\Phi)/\partial(\alpha,\beta)|$
vanishes. This condition is nothing but
Eq.~(\ref{eqn:critical-curve-Jacobian}) characterizing the critical
curve, which divides the parameter space of background solutions into the
high-entropy branch and the low-entropy branch. Therefore,  
\begin{eqnarray}
 \mbox{Critical curve}
  & \Leftrightarrow &
  1/G_\mathrm{eff} = 0, \nonumber\\
 \mbox{High-entropy branch}
  & \Leftrightarrow &
  1/G_\mathrm{eff} > 0, \nonumber\\
 \mbox{Low-entropy branch}
  & \Leftrightarrow &
  1/G_\mathrm{eff} < 0. \label{eqn:global=local}
\end{eqnarray}

Additionally, it is worth mentioning that the effective Newtion's
constant has a connection with thermodynamic quantities on the brane. 
In the 4D case we can define a temperature of the de Sitter space by the
Hubble parameter $H$ as $T = H/2\pi$.
Then, in a straightforward manner, the specific heat is given by 
\begin{equation}
 T\left(\frac{\partial S}{\partial T}\right)
  = - \frac{2\pi}{G_\mathrm N H^2}.
\end{equation}
Now, concerning the 4D de Sitter brane, we consider a specific heat
regarding $T=h/2\pi$ as its temperature.
From Eq.~(\ref{eq:first_law}) we obtain
\begin{equation}
 T\left(\frac{\partial S}{\partial T}\right)_{\widehat\Phi, \sigma_-}
  = \kappa_- h\left(\frac{\partial \widehat S}{\partial h}\right)_{\widehat\Phi}
  = \frac{\pi\kappa_-}{3h^3}\left(\frac{\partial \eta}{\partial h}\right)_{\widehat\Phi}
  = - \frac{2\pi}{G_\mathrm{eff} h^2}.
\end{equation}
This expression is the same as that of the 4D de Sitter
space except for its gravitational ``constant.''
For the 4D de Sitter brane the sign of the specific heat will
change on the critical curve also.
Hence, from the viewpoint of the 4D effective theory, we conclude that
the effective Newton's constant on the brane determines thermodynamic
properties of the braneworld.

\section{Dynamical stability}
\label{sec:dynamical}

In the previous section we have discussed the thermodynamic stability of 
the 6D braneworld model as a property of the background solution.
Now let us investigate the dynamical stability of this spacetime.
For this purpose, we consider linear perturbations around the background
solution.
Perturbations are decomposed into scalar-, vector- and tensor-sectors
according to the representation of the 4D de Sitter symmetry.
If the mass squared of the Kaluza--Klein modes $m^2$ is negative, then 
the corresponding mode is tachyonic and unstable.

Since the background spacetime forms a two-parameter family of solutions, 
the system of the perturbation equations depends on two parameters 
($\alpha$,$\beta$). 
We investigate the lowest mass 
squared $m^2(\alpha,\beta)$ in each sector as a function of the 
two parameters in the square domain $0\leq\alpha\leq 1$, 
$0\leq\beta\leq 1$.
Our strategy for attacking the problem is as follows.
First, we analytically solve the perturbation equations in the
$\alpha = 1$ case for arbitrary $\beta \in [0,1]$, and obtain
$m^2(\alpha=1,\beta)$ on the side $\alpha=1$ of the square domain.
Next, by successively applying the relaxation method, 
we numerically calculate $m^2(\alpha,\beta)$ as $\alpha$ changes 
from $1$ to $0$ for every $\beta$.
In each step of the second procedure, the result of the previous step
with a slightly larger value of $\alpha$ is used as an initial guess for
the relaxation method.

\subsection{Scalar perturbation}

As shown in Appendix~\ref{app:gauge}, the scalar perturbation with 
an appropriate gauge choice is given by
\begin{equation}
 \begin{aligned}
  g_{MN}\mathrm dx^M\mathrm dx^N &= r^2 (1 + \Phi_2 Y)
   g_{\mu\nu}\mathrm dx^\mu\mathrm dx^\nu 
   + [1 + (\Phi_1 + \Phi_2)Y]\frac{\mathrm dr^2}{f}
   + [1 - (\Phi_1 + 3\Phi_2)Y]f\mathrm d\phi^2,\\
  A_M\mathrm dx^M &= (A + a_\phi Y)\mathrm d\phi,
 \end{aligned}
\end{equation}
where $Y$ is the scalar harmonics on the 4D de Sitter space satisfying 
\begin{equation}
 \nabla^2 Y - m^2 Y = 0. 
\end{equation}
Here, $\nabla_\mu$ is the covariant derivative associated with the
4-dimensional de Sitter spacetime with the Hubble $h$.
The Einstein equation and the Maxwell equation are reduced to 
the following two perturbation equations for $\Phi_1$ and $\Phi_2$:
\begin{equation}
 \begin{aligned}
  \Phi''_1 + 2\left(\frac{f'}{f} + \frac{5}{r}\right)\Phi'_1 -
   \frac{4\Lambda_6}{f}(\Phi_1 + \Phi_2)
   + \frac{m^2 + 18h^2}{r^2f}\Phi_1 &= 0,\\
  \Phi''_2 + \frac{4}{r}\Phi'_2 + \frac{m^2}{2r^2f}(\Phi_1 + 2\Phi_2) &= 0,
  \label{eq:scalar_perturbation-original}
 \end{aligned}
\end{equation}
where the prime denotes the derivative with respect to $r$, and $m^2$ is the
eigenvalue of the harmonics.

The boundary conditions at $r = \alpha, 1$ are obtained by 
setting the coefficients of $1/f$ in
(\ref{eq:scalar_perturbation-original}) to zero 
at the brane positions, as in \cite{Yoshiguchi:2005nn}. 
Hence we have
\begin{equation}
 \left. 2 f'\Phi_1' - 4\Lambda_6(\Phi_1+\Phi_2) +
  \frac{m^2+18h^2}{r^2}\Phi_1 \right|_{r=\alpha,1}
 = \left. (\Phi_1+2\Phi_2)\right|_{r=\alpha,1} = 0.
\end{equation}
An alternative and more rigorous derivation of the boundary conditions
is to use the formalism developed in \cite{Sendouda:2006bn}. The result
is 
\begin{equation}
 \left.f\Phi_1\right|_{r=\alpha,1}
  = \left[2f'\Phi_2 + (f\Phi_1)'\right]_{r=\alpha,1} = 0.
  \label{eq:scalar_BC-original}
\end{equation}
These two sets of boundary conditions lead to the same Taylor expansion
of $\Phi_1$ and $\Phi_2$ in the neighborhood of the boundaries and,
thus, are equivalent.

In the original coordinate system $(r, \phi)$ the geometry in the $\alpha=1$ 
limit appears to be singular because two positive roots of $f(r)$ 
corresponding to the brane positions coincide. 
Actually, the proper distance between the $r_\pm$-branes remains finite 
and it turns out that the apparent singularity due to coordinate
artifacts is removed by appropriate coordinate transformations. 
Indeed, in the coordinate system ($w$,$\varphi$) defined by 
\begin{equation}
 w = \frac{2r - (1+\alpha)}{1-\alpha}, \quad
  \varphi = (1-\alpha)\phi,\label{eq:new_coordinates}
\end{equation}
the geometry is obviously regular in the $\alpha=1$ limit. Note that the
coordinate $w$ ranges over the interval $[-1,1]$. 
Since the branes are always located at $w=\pm 1$ for any $\alpha$ (even 
for $\alpha = 1$), the new coordinate $w$ is more useful for numerical
calculation than the original one $r$. 
Therefore we rewrite the perturbation equation in terms of $w$ as
\begin{equation}
 \begin{aligned}
  \partial_w^2\Phi_1 +
  2\left\{\frac{\partial_w\bar{f}}{\bar{f}} +
  \frac{5(1-\alpha)}{[(1-\alpha)w + (1+\alpha)]}\right\}\partial_w\Phi_1 -
   \frac{1}{\bar{f}}(\Phi_1 + \Phi_2)
   + \frac{2\mu^2 + 18h^2/\Lambda_6}{[(1-\alpha)w+(1+\alpha)]^2\bar{f}}\Phi_1 &= 0,\\
  \partial_w^2\Phi_2 + \frac{4(1-\alpha)}{[(1-\alpha)w+(1+\alpha)]}\partial_w\Phi_2 + \frac{\mu^2}{[(1-\alpha)w+(1+\alpha)]^2\bar{f}}(\Phi_1 + 2\Phi_2) &= 0,
  \label{eq:scalar_perturbation}
 \end{aligned}
\end{equation}
where 
\begin{equation}
 \begin{aligned}
  \bar{f}(w)
  \equiv& \frac{f(r)}{\Lambda_6(1-\alpha)^2}\\
  =& \frac{1}{40 (1-\alpha)^2}
  \left\{ -[(1-\alpha)w + (1+\alpha)]^2 +
  \frac{4 (1+\alpha+\alpha^2+\alpha^3+\alpha^4) \beta}
       {1+\alpha+\alpha^2} \right.\\
   & \left. +
  \frac{32 (1+\alpha) [1+\alpha^2 (1+\alpha^2+\alpha^4)
  (1-\beta)-(1+\alpha^3) \beta]}
       {[(1-\alpha)w + (1+\alpha)]^3 (1+\alpha+\alpha^2)} -
  \frac{256 \alpha^3 (1+\alpha+\alpha^2+\alpha^3+\alpha^4) (1-\beta)}
       {[(1-\alpha)w + (1+\alpha)]^6 (1+\alpha+\alpha^2)} \right\}
 \end{aligned}
\end{equation}
and
\begin{equation}
 \mu^2 = \frac{m^2}{2\Lambda_6}.
\end{equation}
The boundary condition is written as
\begin{equation}
 \begin{aligned}
  \left. 2 \partial_w\bar{f}\partial_w\Phi_1 - (\Phi_1+\Phi_2) +
  \frac{2\mu^2+18h^2/\Lambda_6}{[(1-\alpha)w+(1+\alpha)]^2}\Phi_1
  \right|_{w=\pm 1}&=0, \\
  \left. (\Phi_1+2\Phi_2)\right|_{w=\pm 1} &= 0.
 \end{aligned}
 \label{eq:scalar_BC}
\end{equation}

\subsubsection{Analytic solution for $\alpha=1$}

Now we show the analytic solution for $\alpha = 1$.
In this case the geometry of the extra dimensions becomes locally a round
2-sphere as in Ref.~\cite{Freund:1980xh} and the warp factor is a
constant. Hence, the background spacetime for $\alpha = 1$ corresponds
to the football-shaped extra
dimensions~\cite{Carroll:2003db,Garriga:2004tq}.

In the $\alpha = 1$ limit, the perturbation
equations (\ref{eq:scalar_perturbation}) are reduced to 
\begin{equation}
 \begin{aligned}
  (1-w^2) \partial_w^2 \Phi_1 - 4w \partial_w\Phi_1
  - \left(2 - \frac{4\mu^2}{4-3\beta}\right) \Phi_1
  - \frac{8}{4-3\beta}\Phi_2
  &= 0,\\
  (1-w^2)\partial_w^2\Phi_2 + \frac{2\mu^2}{4-3\beta} (\Phi_1 + 2\Phi_2)
  &=0.
 \end{aligned}
\end{equation}
and the boundary conditions are given by
\begin{equation}
  \left.-2w(4 - 3\beta)\Phi'_1 - 4\Phi_2
  + (- 4 + 3\beta + 2\mu^2)\Phi_1\right|_{w=\pm 1} = 0,\quad
  \left.\Phi_1 + 2\Phi_2\right|_{w=\pm 1} = 0.
\end{equation}
The solution satisfying these boundary conditions is 
\begin{equation}
 \Phi_1^{(\pm)} = -2P_n(w) + \frac{4-3\beta}{2\mu^2_{(\pm)}}[n(n+1)P_n(w) -
  2wP'_n(w)],
 \quad \Phi_2^{(\pm)} = P_n(w),
\end{equation}
up to an overall constant, 
where 
$n=0,1,2,\cdots$, and 
$P_n$ is the Legendre function of the first kind.
We obtain 
the mass spectrum of the scalar perturbation as
\begin{equation}
 \mu^2 = 
 \mu_{(\pm)}^2
 \equiv
 \frac{2-3\beta+n (1+n) (4-3\beta) \pm 
       \sqrt{(2-3\beta)^2+12 n (1+n) (1-\beta) (4-3\beta)}}
      {4}.
\end{equation}

The mass of the lowest mode is given by
\begin{equation}
 \mu^2_{(+)} (n=0) = 1 - \frac{3}{2}\beta.
\end{equation}
Thus, for $\beta \ge 2/3$, the lowest 
$\mu^2$ becomes negative and the background spacetime is
destabilized\cite{Contaldi:2004hr,Bousso:2002fi,Martin:2004wp}.

\subsubsection{The lowest mass of the KK mode}

Now we numerically solve the perturbation equations and obtain the
lowest KK mass of the scalar mode for general $\alpha$.
Using the analytic solution for $\alpha = 1$ 
as a first initial guess, we solve the problem for a slightly 
smaller value of $\alpha$ by the relaxation method.
Then, as we change $\alpha$ to smaller values towards $0$ step by step, 
we in turn use the numerical solution of the previous step as an initial
guess for the relaxation method and obtain the mass squared for each
$\alpha$ and $\beta$.

\begin{figure}[t]
 \begin{center}
  \includegraphics[width=.6\linewidth]{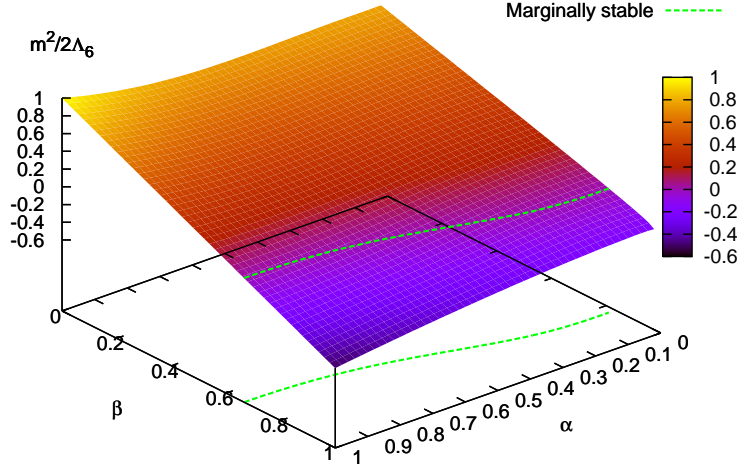}
  \caption{$m^2$ for the lowest mode of the scalar perturbation. When
  $\beta=0$ the 4D spacetime is the Minkowski. As $\beta$ becomes larger
  (i.e.\ the 4D Hubble becomes larger), $m^2$ decreases and eventually
  becomes negative beyond some value of $\beta$. The line represents the
  boundary between the stable and the unstable region, where the
  massless modes appear.}
  \label{fig:scalar_KKmass}
 \end{center}
\end{figure}

Fig.~\ref{fig:scalar_KKmass} shows $m^2$ for the lowest mode of the scalar
perturbation.
For $\beta=0$, which means the 4D spacetime is the Minkowski, we have
reproduced the result of our previous study~\cite{Yoshiguchi:2005nn} 
that there is no unstable (i.e.\ neither massless nor tachyonic)
mode. 
We find that as $\beta$ becomes larger, namely as the Hubble parameter on the
brane becomes larger, $m^2$ decreases and eventually
becomes negative.
Therefore we conclude that when the Hubble on the brane is too large the
extra dimensions are destabilized and such configurations are unstable
for general $\alpha$.
In addition, in the case of vanishing flux ($\beta = 1$) the
configurations are always unstable.

\subsubsection{The threshold massless mode}

Supported by the result shown in Fig.~\ref{fig:scalar_KKmass}, the
assumption that $m^2$ is a continuous function of $\alpha$ and
$\beta$ is made. Hence, it must vanish on the boundary between the stable and
unstable regions. In other words, there should appear a massless mode as
one moves from the stable/unstable region to the unstable/stable
region.

We can actually obtain this threshold massless mode analytically.
The general solution of (\ref{eq:scalar_perturbation-original}) with
$m^2=0$ is given by
\begin{equation}
 \Phi_1 = \frac{1}{f}\left(\frac{c_1}{r^3} + \frac{c_2}{r^6} + c_3 r^2 +
		     \frac{c_4}{r}\right),\quad
  \Phi_2 = \frac{5}{2\Lambda_6}\left(4c_3 + \frac{c_4}{r^3}\right),
\end{equation}
where $c_i (i=1,2,3,4)$ are arbitrary constants.
By imposing the boundary conditions (\ref{eq:scalar_BC-original}), we
will be able to obtain the boundary between the stable and the unstable
region on the $(\alpha,\beta)$ plane, where the massless modes appear.
The boundary conditions lead to
\begin{equation}
 \begin{pmatrix}
  \alpha^2 & \alpha^{-1} & \alpha^7 & \alpha^4\\
  1 & 1 & 1 & 1\\
  6\Lambda_6\alpha^6 & 12\Lambda_6\alpha^3 & 
  -20b^2\alpha^3 + 4\Lambda_6\alpha^{11} - 120\mathcal M\alpha^6 &
  -5b^2 + 4\Lambda_6\alpha^8 - 30\mathcal M\alpha^3\\
  6\Lambda_6 & 12\Lambda_6 & -20b^2 + 4\Lambda_6 - 120\mathcal M &
  -5b^2 + 4\Lambda_6 - 30\mathcal M
 \end{pmatrix}
 \begin{pmatrix}
  c_1 \\ c_2 \\ c_3 \\ c_4
 \end{pmatrix}
 =
 \begin{pmatrix}
  0 \\ 0 \\ 0 \\ 0
 \end{pmatrix},
\end{equation}
where $\mathcal M$ and $b$ are functions of $(\alpha, \beta)$ determined
by $f(\alpha) = f(1) = 0$.
In order for this equation to allow a non-trivial solution for 
$(c_1,c_2,c_3,c_4)$, the determinant of the matrix must vanish.
From this condition, we obtain the relation between $(\alpha,\beta)$ as 
\begin{equation}
 \begin{aligned}
  \beta = 
  \beta_\mathrm{cri}(\alpha)
  &\equiv
  \frac{1}
       {12 (1+\alpha+\alpha^2)^2 (1+\alpha+\alpha^2+\alpha^3+\alpha^4)}
  \bigl[11 + 33\alpha + 66\alpha^2 + 85\alpha^3 + 90\alpha^4 + 85\alpha^5 +
   66\alpha^6 + 33\alpha^7 +  11\alpha^8 \\
  &\quad - (1+\alpha) (1+2\alpha+4\alpha^2+2\alpha^3+\alpha^4)
           \sqrt{1 + 34\alpha^3 + \alpha^6}\bigr].
 \end{aligned}
\end{equation}
This relation is exactly identical with (\ref{eq:critical_curve}), which 
defines the critical curve on which the high-entropy branch and the
low-entropy branch merge.
The threshold massless mode divides the whole square
$(\alpha,\beta)$-domain into positive and negative $m^2$ regions, while
the critical curve in Sec.~\ref{sec:thermodynamics} divides the whole
square $(\alpha,\beta)$-domain into high- and low-entropy
regions. Therefore, 
the high-entropy and the low-entropy branches in Fig.~\ref{fig:entropy}
exactly agree with the positive and negative $m^2$ regions in
Fig.~\ref{fig:scalar_KKmass}, respectively.
This fact implies that in this spacetime thermodynamically preferred
configurations are dynamically stable, while thermodynamically
non-preferred ones have tachyonic modes and become dynamically unstable.
The onset of dynamical instability coincides with the onset of
thermodynamic instability.
This is similar to the Gubser--Mitra conjecture for extended black
objects~\cite{Gubser:2000mm,Gubser:2000ec}.

\subsection{Vector perturbation}

As shown in Appendix \ref{app:gauge}, 
the vector perturbation with an appropriate gauge choice is given by
\begin{equation}
 \begin{aligned}
  g_{MN}\mathrm dx^M\mathrm dx^N &= r^2 g_{\mu\nu}\mathrm dx^\mu\mathrm dx^\nu 
   + 2h_{(\mathrm T)\phi}V_{(\mathrm T)\mu}\mathrm dx^\mu\mathrm d\phi
   + \frac{\mathrm dr^2}{f} + f\mathrm d\phi^2,\\
  A_M\mathrm dx^M &= a_{(\mathrm T)}V_{(\mathrm T)\mu}\mathrm dx^\mu + A\mathrm d\phi,
 \end{aligned}
\end{equation}
where $V_{(\mathrm T)\mu}$ is the vector harmonics on the 4D de Sitter space,
which satisfies the transverse condition, $\nabla^\mu V_{(\mathrm T)\mu} = 0$.
The Einstein equation and the Maxwell equation are reduced to two
perturbation equations with respect to $h_{(\mathrm T)\phi}$ and
$a_{(\mathrm T)}$,
\begin{equation}
 \begin{aligned}
  \frac{1}{r^4}\left[r^6\left(
  \frac{h_{(\mathrm T)\phi}}{r^2}\right)'\right]' - 
  \frac{2b}{r^4}a'_{(\mathrm T)} +
  \frac{m^2 + 6h^2}{r^2f}h_{(\mathrm T)\phi} &= 0,\\
  (r^2f a'_{(\mathrm T)})' + b\left(
  \frac{h_{(\mathrm T)\phi}}{r^2}\right)' + m^2 a_{(\mathrm T)} &= 0,
 \end{aligned}
\end{equation}
where 
$m^2$ is the eigenvalue of vector harmonics on the de Sitter space with
the Hubble constant $h$:
\begin{equation}
 \nabla^2 V_{(\mathrm T)\mu} - (m^2 + 3h^2)V_{(\mathrm T)\mu} = 0, \quad
  \nabla^{\mu} V_{(\mathrm T)\mu} = 0.
\end{equation}

It is again convenient to use the coordinates ($w$,$\varphi$) defined in
Eq.~(\ref{eq:new_coordinates}) 
which do not have any singular behaver in the limit $\alpha\to 1$. 
The perturbation equations and the boundary conditions requiring 
regularity at the brane positions are written in
the coordinates ($w$,$\varphi$) as
\begin{equation}
 \begin{aligned}
  \frac{1}{r^4}\partial_w\left[r^6\partial_w\left(\frac{\Psi_2}{r^2}\right)\right] - \frac{1}{r^4}\partial_w\Psi_1 +
  \frac{\mu^2 + 3h^2/\Lambda_6}{2r^2\bar{f}}\Psi_2 &= 0,\\
  \partial_w(r^2\bar{f} \partial_w\Psi_1) + \frac{b^2}{2}\partial_w\left(\frac{\Psi_2}{r^2}\right) + \frac{\mu^2}{2} \Psi_1 &= 0,
 \end{aligned}
\end{equation}
and 
\begin{equation}
 \begin{aligned}
  (\mu^2 + 3h^2/\Lambda_6)\Psi_2|_{w = \pm 1} &= 0,\\
  \left.
  r^2\partial_w\bar{f} \partial_w\Psi_1 +
  \frac{b^2}{2}\partial_w\left(\frac{\Psi_2}{r^2}\right) +
  \frac{\mu^2}{2} \Psi_1\right|_{w = \pm 1} &= 0,
 \end{aligned}
\end{equation}
where
\begin{equation}
 \Psi_1 \equiv b a_{(\mathrm T)}, \quad \Psi_2 \equiv \frac{h_{(\mathrm T)\phi}}{1 - \alpha}.
\end{equation}
Note that the factor $(1-\alpha)^{-1}$ in the definition of $\Psi_2$
reflects the fact that the relation between the original angular
coordinate $\phi$ and the new coordinate $\varphi$ includes the factor
$(1-\alpha)$: 
$h_{(\mathrm T)\phi}\mathrm d\phi = \Psi_2\mathrm d\varphi$. 
On the other hand, the factor $b$ in the definition of $\Psi_1$ is not
essential but is just to absorb the factor $b$ in the perturbation equations.

\subsubsection{Analytic solution for $\alpha=1$}

For $\alpha = 1$ the perturbation equations become
\begin{equation}
 \begin{aligned}
  \partial_w[(1-w^2) \partial_w\Psi_1] + \frac{8(1-\beta)}{4-3\beta}\partial_w\Psi_2
    + \frac{4\mu^2}{4-3\beta}\Psi_1
  &= 0,\\
  (1-w^2) [\partial_w^2\Psi_2-\partial_w\Psi_1]
    + \frac{4\mu^2+2\beta}{4-3\beta}\Psi_2
  &= 0,
 \end{aligned}
\end{equation}
where $\mu^2 = m^2 /2\Lambda_6$.
The solution satisfying the boundary conditions is given by
\begin{equation}
 \begin{aligned}
  \Psi_1^{(\pm)}(w)
  &= \left(\beta \pm
           \sqrt{\beta^2 + 8n (1+n) (1-\beta) (4-3\beta)}\right)P_n(w),\\
  \Psi_2^{(\pm)}(w)
  &= \frac{(4-3\beta) (1-w^2) P_n'(w)}
          {\beta \pm \sqrt{\beta^2+8n (1+n) (1-\beta) (4-3\beta)}},
 \end{aligned}
\end{equation}
up to an overall constant, and the mass spectrum of the vector
perturbation $\mu^2$ is
\begin{equation}
 \mu^2 = 
 \mu^2_{(\pm)} 
 \equiv
 \frac{n (1+n) (4-3\beta)-\beta \pm 
       \sqrt{\beta^2+8 n (1+n) (1-\beta) (4-3\beta)}}
      {4},
\end{equation}
where $n = 0,1,2,\cdots$ for $\mu_{(+)}^2$ and $n=1,2,\cdots$ for
$\mu_{(-)}^2$.
There are two lowest mass modes for $\mu_{(+)}^2(n=0)$ and
$\mu_{(-)}^2(n=1)$, and both are massless.
In other words the vector perturbations have two zero modes:
\begin{equation}
 \Psi_1 = b a_{(\mathrm T)} = c_1 + 8 c_2 (1-\beta)^2 w,\quad
  \Psi_2 = h_{(\mathrm T)\varphi} = c_2 \frac{4-3\beta}{8}(1-w^2),
\end{equation}
where $c_1$ and $c_2$ are arbitrary constants.
As shown below for general $\alpha$, 
these two zero modes represent physical degrees of freedom of 
the Maxwell field and the gravi-photon.

\subsubsection{Zero modes for general $\alpha$}

For general $\alpha$ we find two zero modes corresponding to the above ones as
\begin{equation}
 a_{(\mathrm T)} = c_1 + c_2 A, \quad h_{(\mathrm T)\phi} = c_2 f,
\end{equation}
where $c_1$ and $c_2$ are arbitrary constants.
It is obvious that the degree of freedom represented by $c_1$ originates
from the original 6D Maxwell field and, thus, is associated with the 4D
part of the $U(1)$ gauge symmetry. 
On the other hand, as seen below, the other degree of freedom
represented by $c_2$ is associated with coordinate transformation of the
angular coordinate $\phi$ and, thus, can be regarded as a gravi-photon.

We now see this explicitly. 
Let us consider 
an infinitesimal $U(1)$ gauge transformation represented by a parameter
$\chi(x)$ and an infinitesimal coordinate transformation 
$\phi \to \phi + \xi(x)$,
where $\chi(x)$ and $\xi(x)$ depend only on the 4D coordinates.
Under these, the $\mu$-components of the perturbation of the U(1) gauge
potential and the ($\mu\phi$)-components of the metric perturbation
transform as
\begin{equation}
  \delta A_\mu \to \delta A_\mu + \partial_{\mu}\chi
   + A\partial_\mu\xi,  \quad
 h_{\mu\phi} \to h_{\mu\phi} + f\partial_\mu\xi,
\end{equation}
and all other components are unchanged. 
Thus, if we define two $4$-vectors $a^{(1)}_{\mu}$ and $a^{(2)}_{\mu}$
as
\begin{equation}
 \delta A_{\mu} = a^{(1)}_{\mu} + Aa^{(2)}_{\mu}, \quad h_{\mu\phi} = fa^{(2)}_{\mu},
\end{equation}
then the above transformation law is rewritten as two separate $U(1)$
gauge transformations:
\begin{equation}
 a^{(1)}_{\mu} \to a^{(1)}_{\mu} + \partial_{\mu}\chi, \quad
  a^{(2)}_{\mu} \to a^{(2)}_{\mu} + \partial_{\mu}\xi. 
\end{equation}
Clearly, $c_1$ and $c_2$ above are coefficients of the transverse
components of $a^{(1)}_{\mu}$ and $a^{(2)}_{\mu}$, respectively. 

Thus, any instability will not occur for these two modes.

\subsubsection{Stability and the 1st KK mode for general $\alpha$}

Having seen that the two zero modes of the vector perturbation
remain massless for any $\alpha$, we will now examine the 1st KK modes for
general $\alpha$ numerically.
\begin{figure}[t]
 \begin{center}
  \includegraphics[width=.48\linewidth]{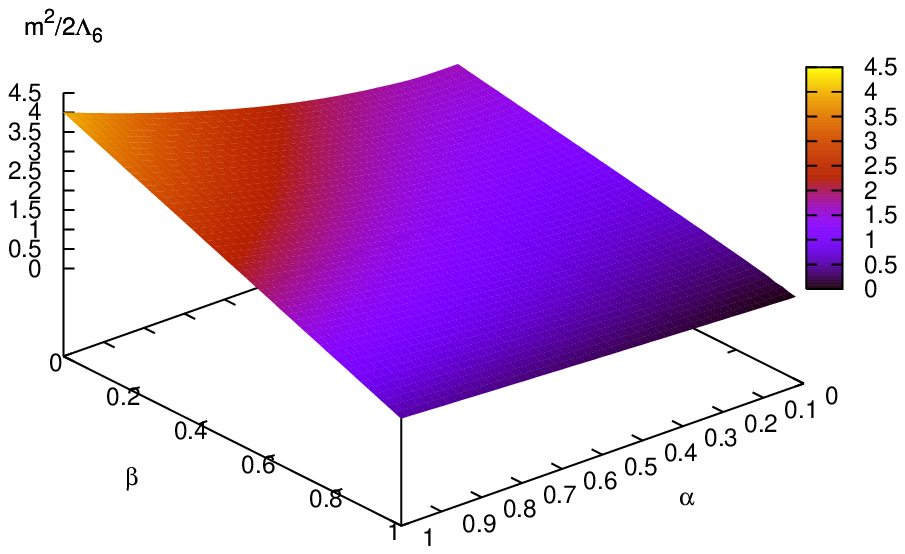}
  \includegraphics[width=.48\linewidth]{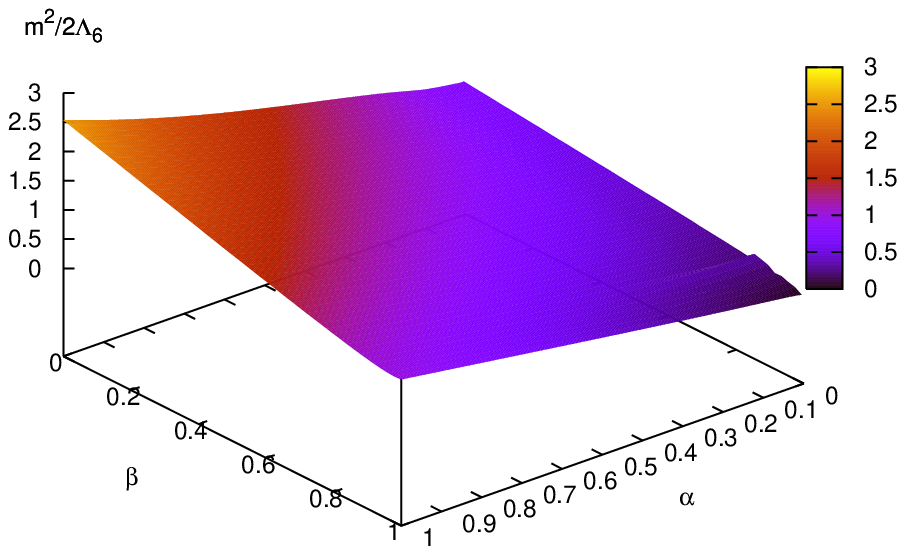}
  \caption{$m^2$ for the first two KK modes of the vector
  perturbation. When $\alpha = 1$, the values correspond to $\mu_{(+)}^2$
  for $n=1$ (left) and $\mu_{(-)}^2$ for $n=2$ (right). These are
  non-negative for the entire region of $(\alpha, \beta)$.}
  \label{fig:vector_KK}
 \end{center}
\end{figure}
Fig.~\ref{fig:vector_KK} shows
$m^2$ for the first two KK modes of the vector perturbation.
We find that $m^2$ is non-negative for the entire region of
$(\alpha,\beta)$.
Therefore, there are no unstable modes in the vector sector.

\subsection{Tensor perturbation}

The tensor perturbation is given by
\begin{equation}
 \begin{aligned}
  g_{MN}\mathrm dx^M\mathrm dx^N &= r^2
  (g_{\mu\nu} + h_{(\mathrm T)}T_{(\mathrm T)\mu\nu})
  \mathrm dx^\mu\mathrm dx^\nu 
  + \frac{\mathrm dr^2}{f} + f\mathrm d\phi^2,\\
  A_M\mathrm dx^M &= A\mathrm d\phi,
 \end{aligned}
\end{equation}
where $T_{(\mathrm T)\mu\nu}$ is the tensor harmonics on the 4D de Sitter space,
which satisfies the transverse and traceless conditions,
$\nabla^\mu T_{(\mathrm T)\mu\nu} = T_{(\mathrm T)}{}^\mu{}_\mu = 0$, 
and
\begin{equation}
 \nabla^2 T_{(\mathrm T)\mu\nu} - (m^2 + 2h^2)T_{(\mathrm T)\mu\nu} = 0.
\end{equation}
The perturbed Einstein equation becomes
\begin{equation}
 \frac{1}{r^2}(r^4fh'_{(\mathrm T)})' + m^2 h_{(\mathrm T)} = 0,
\end{equation}
and there is no relevant equation coming from the Maxwell equation. 
With the coordinate $w$, this is written as
\begin{equation}
 \frac{1}{[(1-\alpha)w+(1+\alpha)]^2}\partial_w\left\{[(1-\alpha)w+(1+\alpha)]^4\bar{f}\partial_w
  h_{(\mathrm T)}\right\} + 2\mu^2 h_{(\mathrm T)} = 0.
\end{equation}

\subsubsection{Analytic solution for $\alpha=1$}

For $\alpha=1$ the perturbation equation becomes
\begin{equation}
 \partial_w[(1 - w^2)\partial_w h_{(\mathrm T)}] + 
  \frac{4\mu^2}{4-3\beta} h_{(\mathrm T)} = 0.
\end{equation}
The regular solution is given by
\begin{equation}
h_{(\mathrm T)} = P_n(w),
\end{equation}
where $P_n$ is the Legendre function of the first kind.
The KK mass squared $\mu^2$ is
\begin{equation}
 \mu^2 = \frac{4 - 3\beta}{4}n(n+1),
\end{equation}
where $n=0,1,2,\cdots$.
It is obvious that the lowest-mass mode is $n=0$ and it is massless,
representing the 4D graviton.

\subsubsection{General $\alpha$}

For the tensor perturbation it is clear that the zero mode is 
$h_{(\mathrm T)} = \text{const.}$, i.e., a homogeneous mode.
\begin{figure}[t]
 \begin{center}
  \includegraphics[width=.5\linewidth]{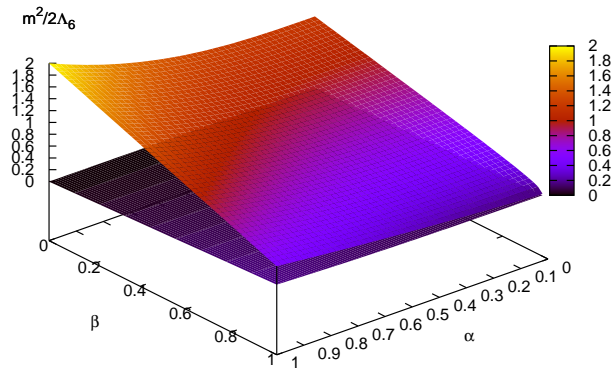}
  \caption{$m^2$ for the first KK mode of the tensor perturbation. The
  lower plane describes the particular mass, $m^2 = 9h^2/4$, for the
  tensor mode on the de Sitter space.}
  \label{fig:tensor_KK}
 \end{center}
\end{figure}
Fig.~\ref{fig:tensor_KK} shows $m^2$
for the first KK mode of the tensor perturbation.
We find that $m^2$ for the KK modes is non-negative\footnote{It is known
that for tensor modes on the de Sitter background the region 
$m^2 < 2h^2$ is forbidden by 
unitarity~\cite{Higuchi:1986py, Deser:2001wx}. Also, the masses in the
region $2h^2 < m^2 < \frac{9}{4}h^2$ are called {\it complementary
series}~\cite{Garidi:2003bg}. Now, the masses for the KK modes 
satisfy $m^2 \ge \frac{9}{4}h^2$.} for the entire region of
$(\alpha,\beta)$.
Therefore, there is no unstable mode in the tensor sector.
(See e.g.~\cite{Hubeny:2002xn}.)

\section{Summary and discussion}
\label{sec:summary}
In this paper we have studied stability of the 6D braneworld
solution with 4D de Sitter branes from two different perspectives.

One is thermodynamic stability of the braneworld solution.
We have defined the de Sitter entropy in the six-dimensional braneworld
by one quarter of the area of the cosmological horizon integrated over
the extra dimensions. We have seen that this definition agrees with the
$4$-dimensional de Sitter entropy defined on each brane. As shown in
Fig.~\ref{fig:entropy}, the de Sitter entropy as a function of the
conserved quantities $\eta$ and $\widehat\Phi$ is not single-valued but
double-valued. Therefore, the de Sitter entropy divides the family of
solutions into two branches, one with higher entropy (the high-entropy
branch) and the other with lower entropy (the low-entropy branch), and
defines the boundary between them (the critical curve).

The other is dynamical stability of the background solutions against
linear perturbations. Perturbations are decomposed into the scalar-,
vector- and tensor-sectors, according to the $4$-dimensional de Sitter
symmetry, and we have calculated the lowest mass squared in each
sector. We have found that when the Hubble expansion rate on the brane
is larger than a critical value, there is a tachyonic mode in the scalar
sector and thus the background is unstable. At the critical value, 
there appears a threshold massless mode in the scalar sector. On the
other hand, there is no unstable mode in the vector- and
tensor-sectors.

We have found that the critical value at which the threshold massless
mode appears is exactly on the critical curve dividing the family of
solutions into the high-entropy branch and the low-entropy
branch. Therefore, we have shown that the low-entropy branch is
dynamically unstable while the high-entropy branch is dynamically
stable. Moreover, we have also seen that the $4$-dimensional effective
Newton's constant is positive in the high-entropy branch and negative in
the low-entropy branch. In summary, we have shown the equivalence of the
following three conditions: 
\begin{itemize}
 \item Thermodynamic stability.
 \item Dynamical stability.
 \item Positivity of the effective Newton's constant.
\end{itemize}
The close connection between thermodynamic stability and dynamical
stability has already been pointed out in the 
literature~\cite{Braden:1990hw,Maeda:1993ap,Gubser:2000mm}. The result
of the present paper, thus, adds yet another example to the list of
systems exhibiting the close connection between thermodynamic and
dynamical properties. In the previous examples in the literature,
however, one has to rely on numerical analysis to show the equivalence
and the boundary between stable and unstable regions is not obtained
rigorously. On the other hand, our system was simple enough to obtain
the boundary analytically. In this sense, we may say that the analysis
in the present paper is the first example in which the equivalence
between thermodynamic stability and dynamical stability was
proved. Moreover, we have shown that for our gravitational system the
thermodynamic and dynamical properties of the spacetime is closely
related to the sign of the effective Newton's constant.

Non-linear evolution of the unstable solutions is one of the important
subjects for the near future. While we have shown that the low-entropy
branch is unstable, our analysis is limited by the linearized 
approximation and we do not know to what final state the spacetime will
evolve from the low-entropy branch. A naive guess is that the system
should evolve to the corresponding solution in the high-entropy branch
with the same values of the conserved quantities, but there remains the
possibility that the system might evolve towards a big crunch
singularity or evolve to another unknown
configuration~\cite{Krishnan:2005su}. Indeed, as seen 
from Fig.~\ref{fig:entropy}, for a solution in the low-entropy branch
with a very large Hubble parameter close to the maximum value, there is
no corresponding solution in the high-entropy branch. However, one can
show that there is a solution with AdS branes for the same conserved
quantities $\eta$ and $\widehat\Phi$. This indicates that the FRW cosmology 
on the brane in this regime should evolve towards a big crunch
singularity unless there is another unknown stable solution without 4D
maximal symmetry. It is worthwhile seeking such solutions without 4D
maximal symmetry and also analyzing properties of solutions with AdS
branes~\cite{DeWolfe:2001nz}. The whole phase structure including the
high- and low-entropy branches, the AdS branch and possibly other
new branches is unexplored.

Inclusion of general matter on the brane is also an interesting subject
as a future work. In the present set-up, the 4D spacetime on the branes
has the maximal symmetry and the stress energy tensor on the brane is 
restricted to the form of vacuum energy, or brane tension. It is
certainly interesting to investigate dynamical stability and
thermodynamic properties of the braneworld with more general Friedmann
universe and matter contents on the brane. For this purpose, we need
some extensions of the model to include arbitrary matter on the
codimension-2 brane~\cite{Peloso:2006cq, Himmetoglu:2006nw,
Papantonopoulos:2006dv, Kobayashi:2007kv}. 
It is also interesting to consider supergravity extensions.

We expect that the close connection between thermodynamic stability
and dynamical stability could be extended to models with branes of higher
codimensions, i.e.\ models with more extra dimensions. We hope to
investigate this subject in the future. Here, as the first step, let us
briefly discuss extensions to models with more extra dimensions but
without branes. We consider general Freund--Rubin flux compactifications
with $\mathrm{dS}_p \times S^q$. The ($p+q$)-dimensional action is 
\begin{equation}
 I = \frac{1}{16\pi}\int \mathrm d^{p+q}x \sqrt{-g}
  \left(R - 2\Lambda - \frac{1}{q!}F_q^2\right),
\end{equation}
where $F_q$ is the $q$-form field for stabilizing the $q$-sphere.
The metric and the $q$-form flux are given by  
\begin{equation}
 \mathrm ds^2 = - \mathrm dt^2 + e^{2ht}\mathrm d\vec{x}_{p-1}^2 +
  \rho^2 \mathrm d\Omega_q^2,
\end{equation}
and
\begin{equation}
 F_q = b \epsilon_{\mu_1 \cdots \mu_q},
\end{equation}
where $\epsilon_{\mu_1 \cdots \mu_q}$ is the volume element of the
$q$-sphere.
The Einstein and Maxwell equations reduce to two algebraic equations:
\begin{equation}
 \begin{aligned}
  (q-1)\rho^{-2} - (p-1)h^2 &= b^2,\\
  (q-1)^2\rho^{-2} + (p-1)^2h^2 &= 2\Lambda.
 \end{aligned}
 \label{eq:algebraic_relations}
\end{equation}
The entropy $S$ is given by 
\begin{equation}
 S = \frac{\Omega_{p-2}}{4h^{p-2}}\Omega_q \rho^q =
  - \frac{\Omega_{p-2}\Phi}{4h^{p-2}b},
\end{equation}
where $\Phi$ is the total flux defined as 
$\Phi \equiv - b\rho^q\Omega_q$ and $\Omega_{p-2,q}$ are respectively the
volume of the unit $(p-2)$-sphere and $q$-sphere.
From (\ref{eq:algebraic_relations}) we can express $S$ as a function
of one parameter.
In this case we can also see that the entropy $S(\Phi)$ as a function
of $\Phi$ is not single-valued
and splits into a high-entropy branch and a low-entropy branch.
The critical point dividing into two branches becomes 
\begin{equation}
 h^2_\mathrm{cri} = \frac{2\Lambda(p-2)}{(p-1)^2(p+q-2)}, \quad
  b^2_\mathrm{cri} = \frac{(p-1)(p+q-2)}{(p-2)(q-1)}h^2_\mathrm{cri},
\end{equation}
which is determined by $\mathrm dS / \mathrm dh = 0$.
These values are nothing but the threshold at which a tachyonic mode
appears in the scalar sector~\cite{Martin:2004wp}. Therefore, we have
shown the equivalence of thermodynamic stability and dynamical
stability in the general Freund--Rubin flux compactifications with
$\mathrm{dS}_p \times S^q$. It is worthwhile investigating a similar
relation in models with branes of codimensions higher than two.

\begin{acknowledgments}
We would like to thank Kei-ichi Maeda for valuable comments.
The work was in part supported by JSPS through a Grant-in-Aid for the
 21st Century COE Program ``Quantum Extreme Systems and Their
 Symmetries'' (S.K.) and a Grant-in-Aid for JSPS Fellows (Y.S.) and by MEXT
 through a Grant-in-Aid for Young Scientists (B) No.~17740134 (S.M.).
\end{acknowledgments}

\appendix
\section{Thermodynamic relations}
\label{sec:thermodynamic_relatios}

In this appendix, we show the thermodynamic relations in the 6D
braneworld solution with the 4D de Sitter space.
The brane positions are determined by the two positive roots of
$f(r)=0$, where
\begin{equation}
f(r) = h^2 - \frac{\Lambda_6}{10}r^2 - \frac{\mathcal M}{r^3} -\frac{b^2}{12r^6}.
\end{equation}
Eliminating $\mathcal M$ from the conditions $f(1) = 0$ and $f(\alpha) = 0$, we obtain
\begin{equation}
 h^2(1 - \alpha^3) - \frac{\Lambda_6}{10}(1 - \alpha^5) -
  \frac{b^2}{12\alpha^3}(\alpha^3 - 1) = 0,
\end{equation}
and this yields
\begin{equation}
\begin{aligned}
\kappa_+ + \kappa_- \alpha^4 &= 
- \frac{1}{2}f'(1) + \frac{1}{2}f'(\alpha)\alpha^4 = 
 \frac{\Lambda_6}{10}(1 - \alpha^5) - \frac{b^2}{4\alpha^3}(\alpha^3 -
 1)\\
 &= h^2(1 - \alpha^3) - \frac{b^2}{3\alpha^3}(\alpha^3 - 1)\\
 &= \frac{3h^4}{4\pi}\mathcal A - b\Phi.
\label{eq:Smarr_relation}
\end{aligned}
\end{equation}
This is the relation between the total area of the horizon
$\mathcal A$, the total magnetic flux of $U(1)$ field $\Phi$, and the proper
periods at the brane positions $\kappa_\pm$ which are respectively
defined as
\begin{equation}
 \begin{aligned}
 \mathcal A
 & = \Delta\phi\int_\alpha^1 r^2 \mathrm dr \frac{4\pi}{h^2}
   = \frac{4\pi}{3h^2}(1 - \alpha^3)\Delta\phi, \\
 \Phi
 & = \int F_{r\phi} \mathrm dr \wedge \mathrm d\phi
   = \frac{b}{3\alpha^3}(\alpha^3 - 1)\Delta\phi,
 \end{aligned}
\end{equation}
and
\begin{equation}
\kappa_+ = -\frac{1}{2}f'(1)\Delta\phi, \quad
\kappa_- = \frac{1}{2}f'(\alpha)\Delta\phi.
\end{equation}
Here $\Delta\phi$ is a given period of the angular coordinate $\phi$ of
extra dimensions.
Recalling Eq.~(\ref{eq:Delta_phi}), we find 
$\kappa_\pm = 2\pi - 8\pi\sigma_\pm$ where $\sigma_\pm$ are tensions of the
$r_\pm$-brane.
We note that Eq.~(\ref{eq:Smarr_relation}) is similar to the
thermodynamic relation for the 6D RNdS black hole.

Furthermore we can obtain the differential relation for the above quantities
which is similar to the first law of black hole mechanics.
The parameters of this spacetime are determined by two equations,
$f(1)=0$ and $f(\alpha)=0$; 
then the variation of parameters in the two equations becomes
\begin{equation}
 \begin{aligned}
  f'(\alpha)\mathrm d\alpha + 2h\mathrm dh - \frac{\mathrm d\mathcal M}{\alpha^3}
  - \frac{b}{6\alpha^6}\mathrm db &= 0,\\
  2h\mathrm dh - \mathrm d\mathcal M - \frac{b}{6}\mathrm db &= 0.
 \end{aligned}
\end{equation}
Eliminating $\mathrm d\mathcal M$, we have
\begin{equation}
 \kappa_- \mathrm d\alpha^4 - \mathcal A \mathrm d\left(\frac{3h^4}{4\pi}\right)
  + \Phi \mathrm db = 0.
\label{eq:differential_relation}
\end{equation}

Although the period of the angular coordinate, $\Delta\phi$, is necessary
for determining the global geometry of the spacetime, locally it is
irrelevant.
In other words, $\mathcal A$, $\Phi$ and $\kappa_\pm$ in
Eq.~(\ref{eq:Smarr_relation}) are ``extensive'' quantities with respect to
the normalization of $\Delta\phi$.
Therefore we express the above relations in terms of
following ``intensive'' quantities, $\eta \equiv \kappa_+/\kappa_-$, 
$\widehat{\mathcal A} \equiv \mathcal A/\kappa_-$, 
and $\widehat\Phi \equiv \Phi/\kappa_-$.
Using these quantities normalized by $\kappa_-$ it is convenient to
discuss the case with the tension of the $r_-$-brane fixed.
From Eqs.~(\ref{eq:Smarr_relation}) and
(\ref{eq:differential_relation}), we have
\begin{equation}
 \eta + \alpha^4 = \frac{3h^4}{4\pi}\widehat{\mathcal A} - b\widehat\Phi,
\end{equation}
and
\begin{equation}
 \mathrm d\alpha^4 - \widehat{\mathcal A} \mathrm d\left(\frac{3h^4}{4\pi}\right)
  + \widehat\Phi \mathrm db = 0.
\end{equation}
Combining these two equations we obtain another expression,
\begin{equation}
 \mathrm d\eta
  = \frac{3h^4}{4\pi}\mathrm d\widehat{\mathcal A} - b\mathrm d\widehat\Phi.
  \label{eq:eta_A_Phi}
\end{equation}
When the tension of the $r_-$-brane is fixed, the variation of $\eta$ is
directly related to the variation of the tension of the $r_+$-brane, i.e.,
$\mathrm d\eta = - \mathrm d\sigma_+$.
In this sense the above expression is important.

\section{Euclidean action}
\label{app:Euclidean_action}

In this appendix, we revisit the derivation of the thermodynamic relations
from the Euclidean action.
(See, for example, \cite{Braden:1990hw}.)
We take an ansatz for the Euclidean 6D metric with topology $S^2 \times S^4$,
\begin{equation}
 \mathrm ds^2 = X(r)^2 \mathrm d\phi^2 + Y(r)^2 \mathrm dr^2 + r^2
  [(1-h^2\rho^2) \mathrm d\tau^2 + (1-h^2\rho^2)^{-1} \mathrm d\rho^2 + \rho^2
  \mathrm d\Omega^2],
\end{equation}
and for the $U(1)$ field,
\begin{equation}
 A_M \mathrm dx^M = A(r)\mathrm d\phi.
\end{equation}
The 4-dimensional part of the metric is a 4-sphere with the radius $h$,
which is the Euclidean continuation of the 4D de Sitter space with the Hubble
parameter $h$ by the Wick rotation of the time coordinate $\tau \to
i\tau$.
The Euclidean time $\tau$ has a period $\Delta\tau = 2\pi/h$ related
with the de Sitter temperature.

We obtain a component of the Einstein tensor, 
\begin{equation}
 G^\phi{}_{\phi} = \frac{2}{r^4}
  \left[r^3
   \left(\frac{1}{Y^2} - h^2\right)
  \right]' ,
\end{equation}
and the non-vanishing components of the energy-momentum tensor are
\begin{equation}
 T^\phi{}_\phi = T^r{}_r = -T^\mu{}_\mu =
  \frac{1}{2}\left(\frac{A'}{XY}\right)^2.
\end{equation}

The nontrivial Maxwell equation is
\begin{equation}
 \left(\frac{r^4A'}{XY}\right)' = 0,
\end{equation}
which is integrated to yield
\begin{equation}
 \frac{r^4A'}{XY} = -b,
\end{equation}
where $b$ is the integration constant.

The component of Einstein equation,  
$G^\phi{}_{\phi} - T^\phi{}_\phi + \Lambda_6 = 0$ is 
\begin{equation}
 \frac{2}{r^4}
  \left[r^3
   \left(\frac{1}{Y^2} - h^2\right)
  \right]' + \Lambda_6 - \frac{b^2}{2r^8} = 0,
\end{equation}
and it is easily integrated to obtain
\begin{equation}
 \frac{1}{Y^2} = h^2 - \frac{\Lambda_6}{10}r^2 - \frac{b^2}{12r^6} -
  \frac{\mathcal M}{r^3},
\end{equation}
where $\mathcal M$ is the integration constant.
The Ricci scalar and the field strength are  
\begin{equation}
 R = -\frac{4}{r^4}\left[r^3\left(\frac{1}{Y^2} -
			     h^2\right)\right]' -
 \frac{2}{r^4XY} \left(\frac{r^4X'}{Y}\right)',
\end{equation}
and
\begin{equation}
 - \frac{1}{2}F_{MN}F^{MN} = - \left(\frac{A'}{XY}\right)^2.
\end{equation}

The Euclidean action is given by
\begin{equation}
 \begin{aligned}
  I_\mathrm E &= -\frac{1}{16\pi}\int \mathrm d^6x_\mathrm E \sqrt{g}
  \left(R - 2\Lambda_6 - \frac{1}{2}F_{MN}F^{MN}\right)\\
  &= - \frac{\Delta\phi}{16\pi}
  4\pi\beta\int_0^{1/h}\!\!\rho^2\mathrm d\rho
  \int_\alpha^1 \mathrm dr 
  \left\{-4XY
  \left[r^3\left(\frac{1}{Y^2} - h^2\right)\right]'
  -
  2 \left(\frac{r^4X'}{Y}\right)' - 2\Lambda_6 r^4XY - \frac{r^4A'^2}{XY}
  \right\}\\
  &= - \frac{\Delta\phi\beta}{6h^3}
  \int_\alpha^1 \mathrm dr 
  \left\{bA' - \left(\frac{r^4X'}{Y}\right)'\right\}\\
  &= -\frac{\beta}{6h^3}(b\Phi + \kappa_+ + \alpha^4\kappa_-),
 \end{aligned}
\end{equation}
where
\begin{equation}
 \left.\Delta\phi\frac{X'}{Y}\right|_{r=\alpha} = \kappa_+ ,\quad
  \left.\Delta\phi\frac{X'}{Y}\right|_{r=1} = -\kappa_- ,
\end{equation}
and
\begin{equation}
 \Phi \equiv \Delta\phi(A|_{r=1} - A|_{r=\alpha}).
\end{equation}
Note that $\kappa_\pm$ and $\Phi$ are the boundary data held fixed in
the current action.
Here, $\beta$ is the inverse of the temperature of the de Sitter
horizon, i.e., $\beta \equiv \Delta\tau$.

Two 3-branes are located at $r = \alpha, 1$, so we require
\begin{equation}
 X(\alpha) = X(1) = 0.
\end{equation}
The regularities at $r = \alpha, 1$ lead to $1/Y^2|_{r=\alpha, 1} = 0$,
and we obtain
\begin{equation}
 h^2 - \frac{\Lambda_6}{10}\alpha^2 - \frac{b^2}{12\alpha^6} -
  \frac{\mathcal M}{\alpha^3}
 = h^2 - \frac{\Lambda_6}{10} - \frac{b^2}{12} - \mathcal M
 = 0.
\end{equation}
From these two conditions we can determine the two integration constants 
$b, \mathcal M$:
\begin{equation}
 \mathcal M(\alpha, h) = (1+\alpha^3)h^2
 - \frac{\Lambda_6 (1-\alpha^8)}{10(1-\alpha^3)},\quad
 b^2(\alpha, h) = \frac{6\Lambda_6 \alpha^3(1-\alpha^5)}{5(1-\alpha^3)} -
 12\alpha^3 h^2,
\end{equation}

The reduced action $I^*$ for fixed $\kappa_\pm$ and $\Phi$ is 
\begin{equation}
 I^* (\Phi, \kappa_+, \kappa_- ; \alpha, h) \equiv 
  - \frac{\pi}{3h^4}(b\Phi + \kappa_+ + \alpha^4\kappa_-).
\end{equation}
Extremizing this action with respect to $\alpha$ and $h$ gives us 
\begin{equation}
 \begin{aligned}
  \frac{\partial I^*}{\partial \alpha} &=
  - \frac{\pi}{3h^4}
  \left(\frac{\partial b}{\partial \alpha}\Phi +
  4\alpha^3\kappa_-\right) = 0,\\
  \frac{\partial I^*}{\partial h} &= 
  \frac{4\pi}{3h^5}(b\Phi + \kappa_+ + \alpha^4\kappa_-) -
  \frac{\pi}{3h^4}\frac{\partial b}{\partial h}\Phi = 0.
 \end{aligned}
\end{equation}
These two conditions are reduced to
\begin{equation}
 \Phi/\kappa_- = 
  4\alpha^3 \left(\frac{\partial b}{\partial \alpha}\right)^{-1}, \quad
  \kappa_+ + \kappa_-\alpha^4 = \frac{3h^4}{\pi} S^* - b\Phi,
  \label{eq:equilibrium_condition}
\end{equation}
where $S^*$ is defined as
\begin{equation}
 S^*(\Phi ; \alpha, h) \equiv 
  \frac{\pi\Phi}{12h^3}\frac{\partial b}{\partial h}
  = - \frac{\pi\alpha^3\Phi}{bh^2}.
\end{equation}
We denote $(\alpha, h)$ satisfying the conditions
(\ref{eq:equilibrium_condition}) for given $\kappa_\pm$ and $\Phi$ as
$\alpha_{\mathrm e}(\Phi, \kappa_+, \kappa_-)$ and 
$h_{\mathrm e}(\Phi, \kappa_+, \kappa_-)$, 
and we have a classical action
\begin{equation}
 \begin{aligned}
  I[\Phi, \kappa_+, \kappa_-]
  &\equiv I^*(\Phi, \kappa_+, \kappa_- ; \alpha_{\mathrm e}, 
  h_{\mathrm e})
  = - \frac{\pi}{3h_{\mathrm e}^4}(b_{\mathrm e}\Phi + \kappa_+ +
  \alpha_{\mathrm e}^4\kappa_-)\\
  &= - S^*(\Phi ; \alpha_{\mathrm e}, h_{\mathrm e})
  \equiv - S[\Phi, \kappa_+, \kappa_-],
 \end{aligned}
\end{equation}
where $b_{\mathrm e} \equiv b(\alpha_{\mathrm e}, h_{\mathrm e})$.
Moreover, the differential relation is given by 
\begin{equation}
 \mathrm dS = - \mathrm dI = \frac{\pi}{3h_{\mathrm e}^4}(b_{\mathrm e} 
  \mathrm d\Phi + \mathrm d\kappa_+ +
  \alpha_{\mathrm e}^4\mathrm d\kappa_-),
\end{equation}
or by using ``intensive'' variables $\widehat S$, $\widehat \Phi$ and
$\eta$ we obtain the alternative expression 
\begin{equation}
 \mathrm d\widehat S = \frac{\pi}{3h^4}(b \mathrm d\widehat\Phi
  + \mathrm d\eta), \quad
  \eta + \alpha^4 = \frac{3h^4}{\pi} \widehat S - b\widehat\Phi.
\end{equation}

From the conditions (\ref{eq:equilibrium_condition}) we obtain the
stationary points of $I^*$, in other words, a series of classical
equilibrium solutions.
To see the (thermodynamic) stability of the solutions, we have to know
which of the equilibrium solutions has the minimum of $I$ (or the maximum of
$S$) for given $\Phi$ and $\kappa_\pm$.
For example, requiring one of the conditions, 
$\partial I^* /\partial\alpha = 0$, we regard $I^*$ as an one-parameter
function of $h$.
(Note that $\Phi$ and $\kappa_\pm$ are fixed.)
Then, we have 
\begin{equation}
 \begin{aligned}
 \frac{\mathrm dI^*}{\mathrm dh} &= \frac{\partial I^*}{\partial h}
  + \frac{\partial I^*}{\partial \alpha}
  \left(\frac{\partial \alpha}{\partial h}\right)_{\Phi,\kappa_-}
  = \frac{\partial I^*}{\partial h},\\
  \frac{\mathrm d^2I^*}{\mathrm dh^2} &= 
  \frac{\partial^2 I^*}{\partial h^2}  
  + \frac{\partial^2 I^*}{\partial h\partial\alpha}
  \left(\frac{\partial \alpha}{\partial h}\right)_{\Phi,\kappa_-}.
 \end{aligned}
\end{equation}
At a stationary point they become 
\begin{equation}
 \begin{aligned}
  \left.\frac{\mathrm dI^*}{\mathrm dh}
  \right|_{\alpha_{\mathrm e},h_{\mathrm e}} &= 0,\\
  \left.\frac{\mathrm d^2I^*}{\mathrm dh^2}
  \right|_{\alpha_{\mathrm e},h_{\mathrm e}} &= 
  -\frac{4}{h}
  \left[
  \frac{\partial S^*}{\partial h}
  +
  \frac{\partial S^*}{\partial\alpha}
  \left(\frac{\partial \alpha}{\partial h}\right)_{\Phi,\kappa_-}
  \right]
  = -\frac{4}{h}\left(\frac{\partial S}{\partial h}\right)_{\Phi,\kappa_-}.
 \end{aligned}
\end{equation}
Hence, the condition for taking a minimum value of $I$ (a maximum value of $S$)
for fixed $\Phi$ and $\kappa_\pm$ is given by 
\begin{equation}
 \left(\frac{\partial S}{\partial h}\right)_{\Phi,\kappa_-} < 0.
\end{equation}

\section{gauge choice}
\label{app:gauge}

\subsection{Harmonics on $\mathrm{dS}_4$}
We give definitions of scalar, vector, and tensor harmonics on the 4D de
Sitter space.
$g_{\mu\nu}$ is the metric of the 4D de Sitter space with the Hubble
parameter $h$, namely, $R_{\mu\nu\alpha\beta} = h^2
[g_{\mu\alpha}g_{\nu\beta} - g_{\mu\beta}g_{\nu\alpha}]$, and $\nabla_\mu$
is the covariant derivative associated with $g_{\mu\nu}$.

The scalar harmonics $Y$ satisfies 
\begin{equation}
 \nabla^2 Y = m^2 Y.
\end{equation}

The vector harmonics $V_{(\mathrm T)\mu}$ satisfies
\begin{equation}
 [\nabla^2 - 3h^2]V_{(\mathrm T)\mu} = m^2 V_{(\mathrm T)\mu},\quad
  \nabla^\mu V_{(\mathrm T)\mu}
  = 0, 
\end{equation}
and $V_{(\mathrm L)\mu}$ is defined as
\begin{equation}
 V_{(\mathrm L)\mu} = \partial_\mu Y.
\end{equation}

The tensor harmonics $h_{(\mathrm T)\mu\nu}$ satisfies
\begin{equation}
 [\nabla^2 - 2h^2]T_{(\mathrm T)\mu\nu} = m^2 T_{(\mathrm T)\mu\nu}, \quad
  \nabla^\mu T_{(\mathrm T)\mu\nu} = T_{(\mathrm T)}{}^\mu{}_\mu = 0,
\end{equation}
and $T_{(\mathrm{TL,LL,Y})\mu\nu}$ is defined by 
\begin{equation}
 \begin{aligned}
  T_{(\mathrm{TL})\mu\nu} &\equiv \nabla_\mu V_{(\mathrm T)\nu} +
  \nabla_\nu V_{(\mathrm T)\mu},\\
  T_{(\mathrm{LL})\mu\nu} &\equiv \nabla_\mu V_{(\mathrm L)\nu} +
  \nabla_\nu V_{(\mathrm L)\mu} -
  \frac{1}{2}g_{\mu\nu}\nabla^\alpha V_{(\mathrm L)\alpha}\\
  &= 2\nabla_\mu \nabla_\nu Y -
  \frac{m^2}{2}g_{\mu\nu}Y ,\\
  T_{(\mathrm Y)\mu\nu}&\equiv g_{\mu\nu} Y.
 \end{aligned}
\end{equation}

\subsection{Gauge choice}

We expand the perturbed metric in harmonics of the 4D de Sitter space
with the Hubble parameter $h$ as
\begin{equation}
 \begin{aligned}
  \delta g_{MN}\mathrm dx^M\mathrm dx^N =& h_{ab}Y\mathrm dx^a\mathrm dx^b\\
  &+ 2(h_{(\mathrm T)a}V_{(\mathrm T)\mu} + h_{(\mathrm L)a}V_{(\mathrm L)\mu})\mathrm dx^a\mathrm dx^\mu\\
  &+ (h_{(\mathrm T)}T_{(\mathrm T)\mu\nu} + h_{(\mathrm{TL})}T_{(\mathrm{TL})\mu\nu} +
  h_{(\mathrm{LL})}T_{(\mathrm{LL})\mu\nu} + h_{(\mathrm Y)}T_{(\mathrm Y)\mu\nu})\mathrm dx^\mu\mathrm dx^\nu,
 \end{aligned}
\end{equation}
where $Y$, $V_\mathrm{(T,L)}$, and $T_\mathrm{(T,TL,LL,Y)}$ are scalar,
vector, and tensor harmonics, respectively.
Greek indices $\mu, \nu, \cdots$ run over the $4$D de Sitter space and
Latin indices $a, b, \cdots$ run over the $2$D extra dimensions $(r,\phi)$.
The coefficients $h_{ab}$, $h_{(\mathrm T)a}$, $h_{(\mathrm L)a}$,
$h_{(\mathrm T)}$, $h_{(\mathrm{TL})}$, $h_{(\mathrm{LL})}$, and
$h_{(\mathrm Y)}$ depend only on $r$ assuming that the perturbations are
axisymmetric.
The perturbations of the $U(1)$ gauge field can be expanded as
\begin{equation}
 \delta A_M\mathrm dx^M = a_a Y\mathrm dx^a +
  (a_{(\mathrm T)}V_{(\mathrm T)\mu} +
  a_{(\mathrm L)}V_{(\mathrm L)\mu})\mathrm dx^\mu,
\end{equation}

The infinitesimal coordinate transformation and $U(1)$ gauge
transformation are 
\begin{equation}
 \delta g_{MN} \to \delta g_{MN} - \mathcal L_\xi g_{MN},\quad
  \delta A_M \to \delta A_M + \partial_M \chi - \mathcal L_\xi A_{M}.
\end{equation}
Also, the gauge parameters can be expanded as 
\begin{equation}
 \xi_M\mathrm dx^M = (\xi_{(\mathrm T)}V_{(\mathrm T)\mu} +
  \xi_{(\mathrm L)}V_{(\mathrm L)\mu})\mathrm dx^\mu + \xi_a Y\mathrm
  dx^a, \quad \chi = \chi_{(\mathrm Y)} Y.
\end{equation}

The tensor-type component is gauge invariant:
\begin{equation}
 h_\mathrm{(T)} \to \bar h_\mathrm{(T)} = h_\mathrm{(T)}.
\end{equation}

The vector-type components transform as
\begin{equation}
 \begin{aligned}
  h_\mathrm{(TL)} &\to \bar h_\mathrm{(TL)} = h_\mathrm{(TL)} -
  \xi_{(\mathrm T)}, & h_{(\mathrm T)r} &\to \bar h_{(\mathrm T)r} =
  h_{(\mathrm T)r} - \xi'_{(\mathrm T)} + \frac{2}{r}\xi_{(\mathrm T)},\\
  h_{(\mathrm T)\phi} &\to \bar h_{(\mathrm T)\phi} = 
  h_{(\mathrm T)\phi}, & a_{(\mathrm T)} &\to \bar a_{(\mathrm T)} =
  a_{(\mathrm T)},
 \end{aligned}
\end{equation}
where the prime denotes the derivative with respect to $r$.
Choosing $\xi_{(\mathrm T)} = h_\mathrm{(TL)}$, we have 
$\bar h_\mathrm{(TL)} = 0$.
With this gauge choice we use new variables.
Using a part of perturbed Einstein equation we obtain algebraically
\begin{equation}
 h_{(\mathrm T)r} = \frac{C}{r^2f},
\end{equation}
where $C$ is an arbitrary constant for $m^2 = -6h^2$, or $C = 0$ for
$m^2 \neq -6h^2$.
When $m^2 = -6h^2$, this constant corresponds to the pure gauge
mode for tensor perturbations because 
$\nabla^\mu T_{(\mathrm{TL})\mu\nu}
 = [\nabla^2 + 3h^2]V_{(\mathrm T)\nu} = 0$ and 
$T_{(\mathrm{TL})}{}^\mu{}_\mu = 2\nabla^\mu V_{(\mathrm T)\mu} = 0$.
Hence we can choose $C = 0$.

The scalar-type components transform as
\begin{equation}
 \begin{aligned}
  h_{(\mathrm{LL})} &\to \bar h_{(\mathrm{LL})} = h_{(\mathrm{LL})} -
  \xi_{(\mathrm L)}, & h_{(\mathrm Y)} &\to \bar h_{(\mathrm Y)} =
  h_{(\mathrm Y)} - \frac{m^2}{2}\xi_{(\mathrm L)} -2rf\xi_r, \\
  h_{(\mathrm L)r} &\to \bar h_{(\mathrm L)r} = h_{(\mathrm L)r} -
  \xi'_{(\mathrm L)} + \frac{2}{r}\xi_{(\mathrm L)} - \xi_r, & 
  h_{(\mathrm L)\phi} &\to \bar h_{(\mathrm L)\phi} = h_{(\mathrm L)\phi} -
  \xi_\phi,\\
  h_{rr} &\to \bar h_{rr} = h_{rr} - 2\xi'_r - \frac{f'}{f}\xi_r, &
  h_{r\phi} &\to \bar h_{r\phi} = h_{r\phi} - \xi'_\phi +
  \frac{f'}{f}\xi_\phi, \\
  h_{\phi\phi} &\to \bar h_{\phi\phi} = h_{\phi\phi} - ff'\xi_r, &
  a_{(\mathrm L)} &\to \bar a_{(\mathrm L)} = a_{(\mathrm L)} +
  \chi_{(\mathrm Y)} -
  A\frac{\xi_\phi}{f}, \\
  a_r &\to \bar a_r = a_r + \chi'_{(\mathrm Y)} -
  A\left(\frac{\xi_\phi}{f}\right)' ,& 
  a_\phi &\to \bar a_\phi = a_\phi - A'f\xi_r.
 \end{aligned}
\end{equation}
Choosing 
\begin{equation}
 \begin{aligned}
  \xi_{(\mathrm L)} &= h_{(\mathrm{LL})}, & \xi_r &= h_{(\mathrm L)r} -
  h'_{(\mathrm{LL})} + \frac{2}{r}h_{(\mathrm{LL})}, \\
  \xi_\phi &= f\int^r_{C'}\mathrm d\rho \frac{h_{r\phi}(\rho)}{f(\rho)}, &
  \chi_{(\mathrm Y)} &=
  - a_{(\mathrm L)} + A\frac{\xi_\phi}{f},
 \end{aligned} 
\end{equation}
where $C'$ is an arbitrary constant, we have $\bar h_{(\mathrm{LL})} =
\bar h_{(\mathrm L)r} = \bar h_{r\phi} = \bar a_{(\mathrm L)} = 0$.
With this gauge choice we redefine each component as follows:
\begin{equation}
 \begin{aligned}
  \bar h_{(\mathrm Y)} &\equiv \Psi, & \bar h_{(\mathrm L)\phi} &\equiv
  h_{(\mathrm L)\phi}, & \bar h_{rr} &\equiv (\Phi_1 + \Phi_2)/f, \\
  \bar h_{\phi\phi} &\equiv -(\Phi_1 + 3\Phi_2)f, & \bar a_r &\equiv
  a_r, & \bar a_\phi &\equiv a_\phi.
 \end{aligned}
\end{equation}
Moreover, using parts of the perturbed Einstein and Maxwell equations we
algebraically obtain the following relations for four variables:
\begin{equation}
 \Psi = \Phi_2, \quad h_{(\mathrm L)\phi} = Cf, \quad a_r = -
  \frac{A'}{f}h_{(\mathrm L)\phi} = - CA', \quad a_\phi =
  \frac{1}{A'}\left[f'\Phi_2 + \frac{1}{2r^2}(fr^2\Phi_1)'\right],
\end{equation}
where $C$ is an arbitrary constant for $m^2 = 0$, or $C=0$ for 
$m^2\neq 0$.
This constant corresponds to the residual (homogeneous) gauge
freedom, $\xi_\phi = Cf$ and $\chi_{(\mathrm Y)} = CA$.
We can set $C = 0$, i.e., $h_{(\mathrm L)\phi} = a_r = 0$.


\end{document}